
\newbox\leftpage \newdimen\fullhsize \newdimen\hstitle \newdimen\hsbody
\tolerance=10000\hfuzz=10000pt
\def\printertype{ps: }
\def\qms{\def\printertype{qms: }
\ifx\answ\bigans\else\voffset=-.4truein\hoffset=.125truein\fi}
\def\bigans{b }
\message{ big or little (b/l)? }\read-1 to\answ
\ifx\answ\bigans\message{(This will come out unreduced.}
\magnification=1200\baselineskip=14pt plus 2pt minus 1pt
\hsbody=\hsize \hstitle=\hsize 
\else\message{(This will be reduced.} \let\lr=L
\magnification=1000\baselineskip=16pt plus 2pt minus 1pt
\voffset=-.31truein\vsize=7truein\hoffset=-.59truein
\hstitle=8truein\hsbody=4.75truein\fullhsize=10truein\hsize=\hsbody
\output={\ifnum\pageno=0 
  \shipout\vbox{\special{\printertype landscape}\makeheadline
    \hbox to \fullhsize{\hfill\pagebody\hfill}}\advancepageno
  \else
  \almostshipout{\leftline{\vbox{\pagebody\makefootline}}}\advancepageno
  \fi}
\def\almostshipout#1{\if L\lr \count1=1 \message{[\the\count0.\the\count1]}
      \global\setbox\leftpage=#1 \global\let\lr=R
  \else \count1=2
    \shipout\vbox{\special{\printertype landscape}
      \hbox to\fullhsize{\box\leftpage\hfil#1}}  \global\let\lr=L\fi}
\fi

\input mssymb.tex
\font\bigfont=cmr10 scaled\magstep3

\def\section#1#2{\vskip32pt plus4pt \goodbreak \noindent{\bf#1. #2}
	\xdef\currentsec{#1} \global\eqnum=0 \global\thmnum=0}

\newcount\thmnum
\global\thmnum=0
\def\prop#1#2{\global\advance\thmnum by 1
	\xdef#1{Proposition \currentsec.\the\thmnum}
	\bigbreak\noindent{\bf Proposition \currentsec.\the\thmnum.}
	{\it#2} }
\def\define#1#2{\global\advance\thmnum by 1
	\xdef#1{Definition \currentsec.\the\thmnum}
	\bigbreak\noindent{\bf Definition \currentsec.\the\thmnum.}
	{\it#2} }
\def\lemma#1#2{\global\advance\thmnum by 1
	\xdef#1{Lemma \currentsec.\the\thmnum}
	\bigbreak\noindent{\bf Lemma \currentsec.\the\thmnum.}
	{\it#2}}
\def\thm#1#2{\global\advance\thmnum by 1
	\xdef#1{Theorem \currentsec.\the\thmnum}
	\bigbreak\noindent{\bf Theorem \currentsec.\the\thmnum.}
	{\it#2} }
\def\cor#1#2{\global\advance\thmnum by 1
	\xdef#1{Corollary \currentsec.\the\thmnum}
	\bigbreak\noindent{\bf Corollary \currentsec.\the\thmnum.}
	{\it#2} }

\newcount\eqnum
\global\eqnum=0
\def\num{\global\advance\eqnum by 1
	\eqno({\rm\currentsec}.\the\eqnum)}
\def\eqalignnum{\global\advance\eqnum by 1
	({\rm\currentsec}.\the\eqnum)}
\def\ref#1{\num  \xdef#1{(\currentsec.\the\eqnum)}}
\def\eqalignref#1{\eqalignnum  \xdef#1{(\currentsec.\the\eqnum)}}

\def\title#1{\centerline{\bf\bigfont#1}}

\newcount\subnum
\def\Alph#1{\ifcase#1\or A\or B\or C\or D\or E\or F\or G\or H\fi}
\def\subsec{\global\advance\subnum by 1
	\vskip12pt plus4pt \goodbreak \noindent
	{\bf \currentsec.\Alph\subnum.}  }
\def\newsubsec{\global\subnum=1 \vskip6pt\noindent
	{\bf \currentsec.\Alph\subnum.}  }
\def\today{\ifcase\month\or January\or February\or March\or
	April\or May\or June\or July\or August\or September\or
	October\or November\or December\fi\space\number\day,
	\number\year}

\def\intsec{I}
\def\stoesec{II}
\def\sslasec{III}
\def\scdsec{IV}
\def\tdpssec{V}
\def\proofsec{VI}
\def\qntsec{VII}
\def\possec{VIII}
\def\qdefsec{IX}
\def\hil{{\cal H}_r({\cal D})}
\def\bC{\Bbb{C}}
\def\bR{\Bbb{R}}
\def\bZ{\Bbb{Z}}
\def\scd{{\cal D}}
\def\aut{{\rm Aut}({\cal D})}
\def\cd{D}
\def\co{{\cal O}}
\def\smooth#1{C^\infty(#1)}
\def\dmu#1{d\mu_r(#1)}
\def\rker#1#2{K^r(#1,#2)}
\def\star{^\ast}
\def\ssub#1{_{\scriptscriptstyle #1}}
\def\del{\partial}
\def\dbr{\overline\partial}
\def\ol{\overline}
\def\norm#1{\Vert #1 \Vert}
\def\inv{^{-1}}
\def\ber{\mathop{\rm Ber}}
\def\gz{{\gamma\ssub Z}}
\def\dgz#1#2{{\gz'(0)_{#1#2}}}
\def\ag{{a_\gamma}}
\def\half{{\textstyle{1\over 2}}}

\def\sd{{\cal D}}
\def\cdi{{D^{\scriptscriptstyle I}_{m,n}}}
\def\cdii{{D^{\scriptscriptstyle II}_n}}
\def\cdiii{{D^{\scriptscriptstyle III}_n}}
\def\scdi{{\sd^{\scriptscriptstyle I}_{m,n|q}}}
\def\scdii{{\sd^{\scriptscriptstyle II}_{n|q}}}
\def\scdiii{{\sd^{\scriptscriptstyle III}_{n|q}}}

\def\sgpi{{SU(m,n|q)}}
\def\inq{I_{n|q}}
\def\tr{\mathop{\rm tr}\nolimits}
\def\sml#1{{\scriptstyle #1}}

{\baselineskip=12pt
\nopagenumbers
\line{\hfill \bf HUTMP B328}
\line{\hfill \bf \today}
\vfill
\title{Matrix Cartan Superdomains,}
\vskip.3cm
\title{Super Toeplitz Operators, and Quantization}
\vskip1in
\centerline{{\bf David Borthwick}$^*$, {\bf Slawomir Klimek}$^{**}$
\footnote{$^1$}{Supported in part by the National Science Foundation
under grant DMS--9206936},}
\vskip.3cm
\centerline{{\bf Andrzej Lesniewski}$^*$\footnote{$^2$}
{Supported in part by the Department of Energy under grant
DE--FG02--88ER25065}, and {\bf Maurizio Rinaldi}$^*$\footnote{$^3$}
{Supported in part by the Consiglio Nazionale delle Ricerche (CNR)}}
\vskip12pt
\centerline{ $^*$Lyman Laboratory of Physics}
\centerline{Harvard University}
\centerline{Cambridge, MA 02138, USA}
\vskip12pt
\centerline{ $^{**}$Department of Mathematics}
\centerline{IUPUI}
\centerline{Indianapolis, IN 46205, USA}
\vskip1in\noindent
{\bf Abstract.}
We present a general theory of non-perturbative quantization of
a class of
hermitian symmetric supermanifolds. The quantization scheme is based
on the notion of a super Toeplitz operator on a suitable
$\bZ_2$-graded Hilbert space of superholomorphic functions. The
quantized supermanifold arises as the $\bC^\ast$-algebra generated
by all such operators. We prove that our quantization framework reproduces
the invariant super Poisson structure on the classical supermanifold
as Planck's constant tends to zero.
\vfill\eject}

\section\intsec{Introduction}

\newsubsec
In this paper we continue our program of non-perturbative quantization of
K\"ahler supermanifolds by means of super Toeplitz operators. This procedure
was first applied in [4] to quantize the hyperbolic unit superdisc
and the flat superspace, and it rested on a $\bZ_2$-graded extension
of the results of [12] and [7]. Our goal here is a similar extension
of the results of [6], where a unified scheme for quantization of
Cartan domains was presented. The significance of Cartan domains
lies in their role in classification of hermitian symmetric spaces
of non-compact type; every (irreducible) such space is equivalent
to a Cartan domain. The Cartan domains fall into four infinite series
(called type I, II, III, and IV domains) as well as two exceptional
cases.  We use the term matrix domains to refer to Cartan domains
of types I--III.  The analysis of [6] relies on the Jordan triple approach to
symmetric domains [15], which provides a unified framework for domains
of all types.

\subsec
The definition of a supermanifold which we adopt in this work is that
of Kostant-Berezin-Leites ([14], [3], [16]), enhanced by the use of the
projective tensor products as in [11]. Recall that a smooth supermanifold
$\cal M$ is a ringed space $(M,\co_M)$, where $M$ is an ordinary smooth
manifold (called the base of $\cal M$), and where $\co_M$ is a sheaf of
supercommutative algebras (over $\bR$) satisfying the following
conditions:
\item{($\ast$)} the quotient sheaf $\co_M/[\co_{M,1} + (\co_{M,1})^2]$,
where $\co_{M,1}$ is the odd part of $\co_M$, is isomorphic to the sheaf
of smooth functions on $M$;
\item{($\ast\ast$)} every point of $M$ has a neighborhood $U$ such that
$$
\co_M|U \cong \smooth U \otimes \bigwedge(E),  \num
$$
where $\bigwedge(E)$ is the Grassmann algebra over a finite dimensional
real vector space $E$.

\noindent
We let $C^\infty(\cal M)$ denote the superalgebra
of global sections of $\co_M$ and refer to its elements as smooth
functions on $\cal M$. The definition of a complex supermanifold is
analogous.
The pair $(n_0|n_1)$, where $n_0=\dim_{\bC} M$, $n_1=\dim_{\bC} E$,
is called the (complex) dimension of $\cal M$.
We equip each $\co_M(U)$ with the usual topology of a Frechet
space. Then $\co_M$ becomes a sheaf of nuclear Frechet algebras. A morphism
in the category of supermanifolds is a pair $(\varphi,\varphi^\#)$ where
$\varphi:M\to N$ is a smooth map of the base manifolds and where
$\varphi^\#:\co_N \to \varphi_\ast \co_M$ is a continuous map of
sheaves of algebras over $N$ ($\varphi_\ast \co_M$ denotes the direct
image of $\co_M$ under $\varphi$). A direct product ${\cal M}\times {\cal N}$
of two supermanifolds is a product object in the category of
supermanifolds.  Clearly, ${\cal M} \times {\cal N} = (M\times N,
\co_M \widehat\otimes_\pi \co_N)$, where $\widehat\otimes_\pi$ is
the completed projective tensor product.

\subsec
In this paper we will be concerned with Poisson supermanifolds,
i.e. supermanifolds for which $C^{\infty}({\cal M})$ is a Poisson
superalgebra ([3], [14]). This means that $C^{\infty}({\cal M})$
is equipped with a bilinear mapping
$$
\{\cdot\; ,\;\cdot\}: C^{\infty}(\cal M)\times C^{\infty}(\cal M)\to
C^{\infty}(\cal M)\; ,\num
$$
called a super Poisson bracket, which satisfies the conditions:
$$
\{f,g\}=(-1)^{p(f)p(g)+1}\{g,f\}\; ,\ref{\sponeref}
$$
$$
(-1)^{p(f)p(h)}\{f,\{g,h\}\}+(-1)^{p(h)p(g)}\{h,\{f,g\}\}+
(-1)^{p(g)p(f)}\{g,\{h,f\}\}=0\; ,\ref{\sptworef}
$$
$$
\{f,gh\}=\{f,g\}h+(-1)^{p(f)p(g)}g\{f,h\}\; ,\ref{\spthreeref}
$$
where $f,g,h\in C^{\infty}(\cal M)$, and where $p(f)\in\{0,1\}$ is the
parity of the (homogeneous) element $f\in C^{\infty}(\cal M)$. Conditions
\sponeref\ and \sptworef\ say that $C^{\infty}(\cal M)$ is a Lie
superalgebra, while condition \spthreeref\ says that the super Poisson
bracket obeys the super Leibniz rule. Poisson supermanifolds arise in
physics as phase spaces for classical systems involving both bosons
and fermions. In the examples discussed in this paper, $\cal M$ is
supersymplectic (in fact, super K\"ahler), i.e. it comes equipped with a
supersymplectic (by which we mean even, closed and non-degenerate) two-form
$\omega$.

\subsec
We plan to present a systematic approach to hermitian symmetric
superspaces elsewhere. Here, we take a more modest point of view and
construct explicitly three infinite series of hermitian supermanifolds
which we call the matrix Cartan superdomains of type I, II, and III.
Their key properties are: (i) the base of a Cartan superdomain of type
I--III is an ordinary Cartan domain of the corresponding type;
(ii) each Cartan superdomain is a homogeneous supermanifold [13],
i.e. it is a quotient
of a Lie supergroup by an appropriate Lie subsupergroup; (iii) the isotropy
supergroup of zero contains circular symmetry. Non-trivial super
versions of the two exceptional domains seem not to exist. On the
other hand, it is likely that a complete list of hermitian symmetric
superspaces will include some ``exotic'' examples without classical
counterparts.

The construction of superdomains in this paper can be extended to
superdomains based on the type IV Cartan domains.  We present this
construction in a separate paper [5].

\subsec
The paper is organized as follows. In Section \stoesec\ we
explain the concept
of a super Toeplitz operator and illustrate it by briefly reviewing the
construction of [4].  Section \sslasec\ contains a brief review
of some facts from super linear algebra.  In Section \scdsec\ we present
the explicit constructions of the matrix superdomains.
In Section \tdpssec\ we describe the super analog of the Jordan triple
determinant and give the corresponding
Poisson structures for the Cartan superdomains. The two main results of
this section, namely Theorems \tdpssec.1 and \tdpssec.2, are proven in
Section \proofsec.  In Section \qntsec\ we define the Bergman spaces of
superholomorphic functions on Cartan superdomains and define the
corresponding super Toeplitz operators. We formulate a number of
technical results and the
two main results of this paper, which are Theorems \qntsec.13 and
\qntsec.14. These
theorems state that the map assigning to a function $f$ the Toeplitz
operator with symbol $f$ is a (non-perturbative) quantization of the
Poisson structure defined in Section \tdpssec.
Section \possec\ contains the proof of the positivity property and some
other technical facts from Section \tdpssec, and
Section \qdefsec\ contains the proofs of
Theorems \qntsec.13 and \qntsec.14.

\bigskip\noindent
{\bf Acknowledgement}. We wish to thank Arthur Jaffe for helpful
discussions and a great deal of encouragement.

\section\stoesec{Super Toeplitz operators}

\newsubsec
A central concept of the present series of papers is that of a super
Toeplitz operator. A super Toeplitz operator is a $\bZ_2$-graded
generalization of a Toeplitz operator and arises in the following context.
Let $\scd=(D, {\cal O}_D)$ be a complex supermanifold whose
base $D$ is a domain in $\bC^N$. We choose global odd generators
$\theta_1,\bar\theta_1,\ldots, \theta_{n_1},\bar\theta_{n_1}$, and
for a function $f\in\smooth\scd$ we write
$$
f(z, \theta_1, \bar\theta_1,\ldots, \theta_{n_1} \bar\theta_{n_1})
 = \sum_{\alpha, \beta} f_{\alpha\beta} (z) \theta^\alpha
\bar\theta^\beta,  \num
$$
where $\alpha$ and $\beta$ are multi-indices, $\theta^\alpha =
\theta_1^{\alpha_1} \ldots \theta_{n_1}^{\alpha_{n_1}}$, and each
$f_{\alpha\beta} \in \smooth\cd$.  The complex conjugation of a
product of elements of $\smooth\scd$ reverses the order:
$$
\ol{fg} := \bar g \bar f = (-1)^{p(f)p(g)} \bar f \bar g.  \num
$$

We call a function $f\in\smooth\scd$ bounded if each of the
components $f_{\alpha\beta}$ together with all its derivatives is bounded.
The subspace of bounded smooth functions on $\scd$ is denoted by
$B^\infty(\scd)\subset C^\infty(\scd)$.
We give $B^\infty(\scd)$ the topology of a Frechet space, which
is defined by the following family of norms:
$$
\norm f_t := \sum_{|\mu|+|\nu|\le t} \sum_{\alpha,\beta}\; \sup_{z\in D}
\bigl| \del_z^\mu \del_{\bar z}^\nu f_{\alpha\beta}(z) \bigr|,
\ref{\normt}
$$
where $t \ge 0$, and $\mu,\nu$ are multi-indices of length $n_0$
with $|\mu| := \mu_1 + \ldots + \mu_{n_0}$. The derivatives
$\del_z^\mu$ are defined in the obvious way.

Let $d\mu$ be a volume form on $\scd$ (a ``measure'') such that $\int_\scd
d\mu =1$. The integral
$$
(f,g):=\int_\scd\ol {f(Z)} g(Z)d\mu(Z)\ref{\sesq}
$$
defines a sesquilinear form on $B^\infty(\scd)$.
Unlike the usual forms of this type,
\sesq\ does not need to be positive definite (in fact,
in the examples that we study it is not positive definite). A function
$f\in\smooth\scd$ is called superholomorphic if $\del_{\bar z_j}f =
\del_{\bar\theta_k}f = 0$, for all $j$ and $k$.  The basic
assumption about the measure $d\mu$ is the following positivity property
(which resembles very much the reflection positivity of Euclidean
field theory and statistical mechanics, see e.g. [9]).

\medskip\noindent
{\it The form \sesq\ defines an inner product on the subspace
${\rm Hol}(\scd)$ of $B^\infty(\scd)$ consisting of
superholomorphic functions.}

\medskip\noindent
We let ${\cal H}(\scd,d\mu)$ denote the ($\bZ_2$-graded) Hilbert
space obtained as the completion with respect to \sesq\
of ${\rm Hol}(\scd)$ and call it
the Bergman space. Let $P: B^\infty(\scd)\rightarrow
{\cal H}(\scd,d\mu)$ be a projection map. For
$f\in B^\infty(\scd)$ and $\phi\in {\cal H}(\scd,d\mu)$
we set
$$
T(f)\phi :=PM(f)\phi\; ,\ref{\toepldef}
$$
where $M(f)$ denotes the operator (on $B^\infty(\scd)$) of
multiplication by $f$. The linear operator $T(f): {\cal H}(\scd,d\mu)
\rightarrow {\cal H}(\scd,d\mu)$ is called a super Toeplitz operator
with symbol $f$.

\subsec
To illustrate the above concepts we briefly review the construction
of super Toeplitz operators arising in the quantization of the simplest
hyperbolic supermanifold, namely the super unit disc (see [4] for
the details and proofs). This construction will be generalized
in Sections IV and V to arbitrary Cartan superdomains. The super unit
disc ${\cal U}\equiv U^{1|1}$ is a $(1|1)$-dimensional complex
supermanifold $(U,{\cal O}_U)$ whose base is the open unit disc
$U = \{z\in \bC: |z|<1\}$. We denote the odd generators of $C^{\infty}
({\cal U})$ by $\theta$ and ${\bar\theta}$.

We will use a collective notation for the generators of $C^\infty({\cal U})$,
namely $Z := (z,\theta)$. Consider now the following measure on ${\cal U}$.
For $r\geq 1$ we set
$$
d\mu_r(Z) := {1\over\pi}(1 - Z{\ol Z})^{r-1} d^2z\; d^2\theta.
\ref{\dmrdef}
$$
where $Z\ol Z :=|z|^2+\theta {\bar\theta}$, $d^2z={i\over 2}
dz\wedge d\bar z$ is the volume form on $U$, and $d^2\theta$
is the Berezin integral with $\int \bar\theta\theta d^2\theta =1$.
Using the expansion
$$
(1-|z|^2-\theta {\bar\theta})^{r-1} = (1-|z|^2)^{r-1}
- (r-1)(1-|z|^2)^{r-2}\theta {\bar \theta},  \ref{\izzttexp}
$$
we compute the total integral
$$
\int_{\cal U} d\mu_r(Z) = {r-1\over\pi} \int_U (1-|z|^2)^{r-2} d^2z
= (r-1)\int_0^1 (1-t)^{r-2} dt = 1,\num
$$
i.e. the measure $d\mu_r$ has mass one. Using
\izzttexp\ it is easy to see that the associated sesquilinear form
\sesq\ is not positive definite. On the other hand, for $\phi$
superholomorphic we can write $\phi(Z) = \phi_0(z) + \phi_1(z)
\theta$, so that for such a function,
$$
(\phi,\phi)={{r-1}\over{\pi}}\int_U|\phi_0(z)|^2(1-|z|^2)^{r-2}d^2z
+{1\over{\pi}}\int_U|\phi_1(z)|^2(1-|z|^2)^{r-1}d^2z,\num
$$
which is clearly positive. The projection map $P$ taking bounded elements
of $C^\infty({\cal U})$ to ${\cal H}({\cal U}, d\mu_r)$ is given by
the integral operator
$$
Pf(Z) := \int_{\cal U} K^r(Z,W) f(W) d\mu_r(W),  \num
$$
where
$$
K^r(Z,W) := (1 - Z\ol W)^{-r}  \num
$$
is the Bergman kernel for ${\cal H}({\cal U}, d\mu_r)$. The super Toeplitz
operator, whose symbol is a bounded function $f\in C^\infty({\cal U})$,
is then defined by
$$
\bigl(T_r(f) \phi\bigr)(Z) := \int_{\cal U}K^r(Z,W) f(W) \phi(W)
d\mu_r(W).  \num
$$

\section\sslasec{Some super linear algebra}

\newsubsec
Because this paper involves a good deal of explicit computations with
both supermatrices and ordinary matrices, we review here our conventions.
These follow those of [3]. We call a matrix with entries in a
supercommuting algebra an ordinary matrix if its entries are purely even.
For ordinary matrices, which will typically be denoted
by lower case Roman letters, we use the standard notations of $\bar a$
and $a^t$ to denote conjugate and transpose.  Matrices with purely
odd entries will be denoted by lower case Greek letters, and conjugation
and transposition will be defined just as for ordinary matrices.
Note, however, that
$$
\ol{\alpha\beta} = - \bar\alpha \bar\beta, \qquad
(\alpha\beta)^t = - \beta^t \alpha^t.  \num
$$
Capital Roman letters will denote supermatrices.
We use $\ast$ to denote the hermitian adjoint for these cases.

An $m|n\times k|l$ supermatrix has the form
$$
A= \bordermatrix{&\sml k&\sml l \cr \sml m&a& \alpha \cr
\sml n& \beta &b\cr}, \num
$$
where $a$ and $b$ are ordinary matrices and $\alpha$ and $\beta$
have purely odd entries.  If $l=0$ we will write $m|n \times k$ for
the dimension, and if $n=0$ the dimension will be $m\times k|l$,
i.e. single dimensions always refer to an even component.
The superanalogs of conjugation and transposition are defined as follows:
$$
A^c := \pmatrix{\bar a& -\bar\alpha\cr \bar\beta&\bar b\cr},\num
$$
$$
A^T := \pmatrix{a^t& \beta^t\cr -\alpha^t& b^t\cr}.  \num
$$
Note that $T^2 \ne 1$.
The hermitian adjoint of a supermatrix is given by $A^\ast :=
(A^c)^T$. We use the same symbol as for ordinary matrices because
the same transformation is performed:
$$
A^\ast = \pmatrix{a\star& \beta\star\cr \alpha\star& b\star\cr}. \num
$$

\subsec
The Berezinian [3] of a square supermatrix is defined by the formula
$$
\ber \pmatrix{a&\alpha \cr \beta& b\cr} := {\det(a -
\alpha b\inv \beta) \over \det b} .  \num
$$
We will often write supermatrices in a nonstandard form:
$$
\gamma = \bordermatrix{&\sml m& \sml{n|q}\cr
\sml m &A & B \cr
\sml{n|q}& C & D\cr},  \num
$$
where $A,B,C$, and $D$ are subsupermatrices.  In this case the Berezinian
is
$$
\ber\gamma = \ber D \;\det (A - BD\inv C). \ref{\berforma}
$$

For convenience we state here a formula for the inverse of a
matrix which we will use frequently.  For any ordinary
matrix or supermatrix in block form, we have
$$
\pmatrix{A&B\cr C&D\cr}\inv =
\pmatrix{(A - BD\inv C)\inv & - A\inv B (D-CA\inv B)\inv \cr
-D\inv C (A - BD\inv C)\inv & (D - CA\inv B)\inv \cr}.  \num
$$
The proof is obvious.

\subsec
We include the following useful technical fact to illustrate the
mechanics of dealing with Berezinians.
\lemma\iwzizw{For an $m \times n|q$ supermatrix $A$ and
an $n|q\times m$ supermatrix $B$, we have
$$
\ber (\inq - BA) = \det (I_m - AB), \num
$$
where $\inq$ denotes the dimension $n|q$ identity supermatrix.}

\medskip\noindent{\it Proof.}
We write $A = (a, \alpha)$ and $B = \pmatrix{b\cr \beta\cr}$.
By definition,
$$
\eqalign{\ber (\inq - BA) &= {\det \bigl( I_n - ba - b\alpha
(I_q - \beta\alpha)\inv \beta a\bigr) \over
\det (I_q - \beta\alpha)} \cr
&= {\det \bigl( I_m - ab - ab\alpha\beta
(I_n - \alpha\beta)\inv) \over
\det (I_q - \beta\alpha)}. \cr } \ref{\beriba}
$$
Because the entries of $\alpha$ and $\beta$ anticommute, we have
$$
\eqalign{\det (I_q - \beta\alpha) &= \exp\Bigl\{ \sum_{l=0}^\infty
{1\over l} \tr (\beta\alpha)^l \Bigr\}\cr
&= \exp\Bigl\{ - \sum_{l=0}^\infty
{1\over l} \tr (\alpha\beta)^l \Bigr\}\cr
&= \det (I_n - \alpha\beta)\inv.\cr} \num
$$
Returning to \beriba, this implies
$$
\eqalign{\ber (\inq - BA) &= \det \Bigl( (I_n - ab)(I_n -
\alpha\beta) - ab\alpha\beta \Bigr) \cr
&= \det(I_n - ba - \alpha\beta) \cr
&= \det(I_n - BA)\inv. \qquad\square\cr} \num
$$

\bigskip
Note that an immediate consequence of \iwzizw\ is that \berforma\
is equivalent to
$$
\ber \gamma = \det A \; \ber (D-CA\inv B). \ref{\berformb}
$$

\section\scdsec{Matrix Cartan superdomains}

\newsubsec
In this section we describe the main objects of our study, namely the matrix
Cartan superdomains. Recall (see e.g. [10], [15]) that all symmetric hermitian
domains fall into four series of classical Cartan domains,
with two exceptional domains.
The first three classes are the matrix domains, which are defined
as follows. In the formulas below, $D$ with suitable decorations denotes
a Cartan domain and ${\rm Aut}(D)$ denotes the Lie group of
biholomorphisms of $D$. The definitions of all the Lie groups
involved can be found in [10], whose notation we follow.

\bigskip\noindent
{\it Type I.}
We let
$$
\eqalign{\cdi &:= \bigl\{ z\in {\rm Mat}_{m,n}(\bC):\; I_n - z\star z >
0\bigr\} \cr
&\cong SU(m,n)\big/S(U(m)\times U(n)).  \cr}\num
$$
The group $SU(m,n)$ acts on $\cdi$ by holomorphic automorphisms in the
following way.  We write $\gamma\in SU(m,n)$ in the block form
$$
\gamma = \bordermatrix{&\sml m&\sml n\cr \sml m&a&b\cr
\sml n&c&d\cr},  \num
$$
where the submatrices $a,b,c$, and $d$ have the dimensions
indicated and satisfy
$$
\eqalign{&a\star a - c\star c = I_m, \cr
&a\star b = c\star d, \cr
&d\star d - b\star b = I_n.  \cr}  \num
$$
The corresponding element of Aut$(\cdi)$ is
$$
\gamma:z\mapsto (az+b)(cz+d)\inv.  \num
$$

\bigskip\noindent
{\it Type II.}
We set
$$
\eqalign{\cdii &:=\bigl\{ z\in {\rm Mat}_{n,n}(\bC):\; z^t = z,\; I_n -
z\star z> 0\bigr\}  \cr
&\cong Sp(n)/U(n).  \cr}  \num
$$
The biholomorphic action of $Sp(n)$ on $\cdii$ is defined as follows.
We write $\gamma\in Sp(n)$ as
$$
\gamma = \bordermatrix{&\sml n&\sml n\cr \sml n&a&b\cr
\sml n&\bar b& \bar a\cr},  \num
$$
where $a,b$ satisfy
$$
\eqalign{&a\star a - b^t\ol b = I_n,  \cr
&a^t \bar b = b\star a.  \cr}\num
$$
Then
$$
\gamma: z \mapsto (az+b)(\bar bz + \bar a)\inv  \num
$$
is the corresponding element of Aut$(\cdii)$.

\bigskip\noindent
{\it Type III.}
Let
$$
\eqalign{\cdiii &:= \bigl\{ z\in {\rm Mat}_{n,n}(\bC):\; z^t = -z,\;
I_n - z\star z>0\bigr\}  \cr
&\cong SO^*(2n)/U(n).  \cr}  \num
$$
The action of $SO^*(2n)$ is defined as follows.  We write
$\gamma\in SO^*(2n)$ as a block matrix,
$$
\gamma = \bordermatrix{&\sml n&\sml n\cr \sml n&a&b\cr
\sml n&-\bar b& \bar a},  \num
$$
with $a,b$ such that
$$
\eqalign{&a\star a - b^t\bar b = I_n, \cr
&a^t\bar b = - b\star a.  \cr}  \num
$$
The corresponding element of Aut$(\cdiii)$ is then
$$
\gamma : z\mapsto (az+b)(-\bar bz + \bar a)\inv.  \num
$$

\subsec
A Cartan superdomain $\scd$ is a supermanifold $(D,{\cal O})$,
where $D$ is an ordinary Cartan domain, and where ${\cal O}$ is
a sheaf of superalgebras on $D$ which will be defined case by case
below. We define the superdomains of types I, II, and III, denoted
below by $\scdi$, $\scdii$, and $\scdiii$, respectively.

\noindent
{\it Type I.} We set
$$
\smooth\scdi := \smooth\cdi \otimes \bigwedge(\bC^{m\times q})\; .
 \num
$$
We organize the standard generators of $\bigwedge(\bC^{m\times q})$
into $m\times q$ matrices $\theta = \{\theta_{ij}\}$ and $\bar\theta =
\{\bar\theta_{ij}\}$, and represent the ``points'' of $\scd$ as the
$m\times n|q$ supermatrices
$$
Z = (z,\theta). \ref{\bigzzth}
$$
The matrix dimension $q$ for the odd components is arbitrary.

We define the supermanifolds $\scdii$ and $\scdiii$ as subsupermanifolds of
the type I superdomains.   This is done by imposing constraints on the
generators of $\smooth{{\cal D}^I_{n,n|q}} $, as follows.

\noindent
{\it Type II.} We impose
$$
z - z^t + \theta\theta^t=0. \ref{\symrel}
$$
The fermionic dimension $q$ is again arbitrary for type II.

\noindent
{\it Type III.} We require
$$
z^t+z-\theta\tau_q\theta^t=0,\ref{\skewsymrel}
$$
where $\tau_q$ is the $q\times q$ matrix
$$
\tau_q :=\left(\matrix{0 & iI_{q/2}\cr -iI_{q/2} & 0\cr}\right)\;
.\ref{\taudef}
$$
Note that $q$ must be even for type III superdomain.

Each of the above superdomains $\scd$ admits an action of a Lie
supergroup ${\rm Aut}(\scd)$ of superholomorphic automorphisms. In
all cases, ${\rm Aut}(\scd)$ is an intersection of an orthosymplectic
supergroup with the supergroup $\sgpi$.  This supergroup is defined as
follows.  Its base manifold is $SU(m,n)\times SU(q)$, and its structure sheaf
is
generated by $\gamma_{jk}$ and $\bar\gamma_{jk}$, $1\le j,k\le m+n+q$,
with the following parity assignments:
$$
p(\gamma_{jk}) = p(\bar\gamma_{jk}) = \cases{0,& if $1 < j,k \le m+n$ or
$m+n< j,k \le m+n+q$, \cr
1,& otherwise,  \cr}  \num
$$
and with the following relations.  We write $\gamma$ as a block
supermatrix
$$
\gamma = \bordermatrix{&\sml m&\sml n&\sml q\cr \sml
m&a&b&\rho\cr
\sml n&c&d& \delta\cr \sml q&\alpha&\beta& e\cr} ,  \ref{\fullgamma}
$$
where $a,b,c,d$, and $e$ are even matrices and $\alpha,\beta, \rho$,
and $\delta$ are odd matrices of the dimensions indicated, and require that
$$
\ber \gamma = 1.  \num
$$
The real structure on $\sgpi$ is defined by setting
$$
\gamma\star = J\gamma\inv J,  \ref{\gjgeq}
$$
where
$$
J = \left( \matrix{I_m & 0 & 0 \cr 0 & -I_n & 0 \cr 0 & 0 & -I_q \cr}
\right). \num
$$
Equation \gjgeq\ is equivalent to the set of relations:
$$
\eqalign{
&a\star a - c\star c - \alpha\star\alpha = I_m,  \cr
&a\star b - c\star d - \alpha\star\beta = 0,  \cr
&a\star \rho - c\star \delta - \alpha\star e = 0,  \cr
&b\star b - d\star d - \beta\star\beta = - I_n,  \cr
&b\star \rho - d\star \delta - \beta\star e = 0,  \cr
&\rho\star \rho - \delta\star \delta - e\star e = - I_q.  \cr}  \num
$$

In view of \bigzzth, we will find it convenient to rewrite \fullgamma\ in the
non-standard form
$$
\gamma = \bordermatrix{&\sml m&\sml{n|q}\cr \sml m&A&B\cr
\sml{n|q}&C&D\cr},  \ref{\nonstand}
$$
where $A = a$, and $B, C$, and $D$ are now supermatrices obeying the
relations
$$
\eqalign{&A\star A - C\star C = I_m,  \cr
&A\star B = C\star D ,\cr
&D\star D - B\star B = \inq.  \cr}\ref{\grelns}
$$

Consider now the morphism $\smooth{\scdi}\to
\smooth{SU(m,n|q)}\widehat\otimes_\pi\smooth{\scdi}$ defined by
$$
\gamma : Z \mapsto Z' := (AZ + B)(CZ+D)\inv,  \ref{\gaction}
$$
where, for simplicity, we have suppressed the tensor product symbols
(writing $AZ$ in place of $A\otimes Z$ and so on).  By the relations
\grelns\ this transformation is equivalent to
$$
\eqalign{\gamma(Z) &= (ZB\star + A\star)\inv (ZD\star + C\star) \cr
&= (zb\star + \theta\rho\star + a\star)\inv (zd\star + \theta\delta\star
+ c\star, z\beta\star + \theta e\star + \alpha\star). \cr} \ref{\zpralt}
$$
Clearly $Z'$ defines a new set of generators for $\smooth\scdi$.

\prop\acti{The above morphism defines a transitive action of
$SU(m,n|q)$ on $\scdi$.  Furthermore,
$$
\scdi \cong  SU(m,n|q)\big/ S(U(m)\times U(n|q)).  \ref{\superquo}
$$}

\noindent{\it Proof.}
The fact that $z\star z < I$ implies that $(ZB\star + A\star)$
is invertible, because $A$ is invertible and the non-nilpotent part
of $ZB\star$ is $zb\star$.  The result follows from the corresponding
property of the underlying Cartan domain.

To prove \superquo, we note that the
isotropy subsupergroup of $0$ consists of supermatrices
$$
\left(\matrix{A&0\cr 0&D\cr}\right),  \num
$$
with
$$
A\star A = I_m,\qquad D\star D = \inq. \qquad\square\num
$$

\subsec
We now turn to the type II case. The Lie supergroup acting
on $\scdii$ is denoted by $Sp(n|q)$ and is defined
as the intersection of $SU(n,n|q)$ with
the orthosymplectic supergroup $SpO(n|q)$. The latter is defined
in terms of supermatrices of the form \fullgamma, where $m=n$.
We require that $\ber
(\gamma)=1$, and
$$
\gamma^T K\gamma=K\; ,\ref{\spo}
$$
where $K$ is the supermatrix
$$
K=\left(\matrix{0 & I_n & 0 \cr -I_n& 0 & 0\cr 0 & 0 & I_q\cr}\right)\;
.\num
$$
Solving the relations \gjgeq\
and \spo\ we write the generators of $Sp(n|q)$ in the form
$$
\gamma=\bordermatrix{&\sml n&\sml n&\sml q\cr
\sml n&a & b & \rho \cr \sml n& {\bar b} & {\bar a} & -{\bar\rho}
\cr
\sml q& \alpha & {\bar\alpha} & e \cr}\; ,\qquad {\bar e}=e\;
,\ref{\spogam}
$$
with the entries satisfying
$$
\eqalign{
& a^t{\bar b}-b^* a+\alpha^t\alpha=0\; ,\cr
& a^t{\bar a}-b^*b+\alpha^t{\bar\alpha}=I_n\; ,\cr
& a^t{\bar\rho}+b^*\rho-\alpha^te=0\; ,\cr
& \rho^t {\bar\rho}-\rho^* \rho+e^te=I_q\; .\cr}\ref{\relspo}
$$

Consider now the morphism $\smooth{\scdii}\to \smooth{Sp(n|q)}
\widehat\otimes_\pi\smooth{\scdii}$ defined by
$$
\gamma : Z \mapsto Z' := (AZ + B)(CZ+D)\inv,  \ref{\gactionii}
$$
where
$$
A:=a,\quad B:=(b,\rho),\quad C:=\left(\matrix{{\bar b}\cr \alpha}\right),
\quad D:=\left(\matrix{{\bar a} & -{\bar\rho}\cr {\bar\alpha} & e\cr}
\right)\; .\ref{\nstdf}
$$

\prop\actii{The above morphism defines a transitive action of
$Sp(n|q)$ on $\scdii$.  Furthermore,
$$
\scdii \cong  Sp(n|q)\big/U(n)\times SO(q).  \ref{\superquoii}
$$}

\noindent{\it Proof.}
Clearly, \gactionii\ is well defined by the same argument as for
\acti.  Recall that the defining relation of $\scdii$ was
$$
z - z^t + \theta\theta^t = 0.  \num
$$
To show that this relation is preserved under the action of $Sp(n|q)$,
we recast it as
$$
(I_n,Z) K \pmatrix{I_n\cr Z^T \cr} = 0. \ref{\recast}
$$
Now, from \zpralt\ we can write
$$
(I_n, Z') = (ZB\star + A\star)\inv (I_n, Z) \gamma\star,  \num
$$
so that
$$
(I_n,Z')K\pmatrix{I_n\cr Z'^T} =
(ZB\star + A\star)\inv (I_n,Z) \gamma\star K \gamma^{\ast^T}
\pmatrix{I_n\cr  Z^T}
\bigl((ZB\star + A\star)\inv\bigr)^T. \num
$$
Taking the adjoint and then transpose of the relation
$\gamma^T K \gamma = K$ gives $\gamma\star K \gamma^{\ast^T} =
K$,
so that \recast\ implies
$$
(I_n, Z')K \pmatrix{I_n\cr Z'^T} = 0.  \num
$$

To prove \superquoii, we note that the isotropy supergroup of $0$ consists
of supermatrices
$$
\gamma=\left(\matrix{a & 0 & 0 \cr 0 & {\bar a} & 0 \cr
0 & 0 & e \cr}\right)\; ,\qquad {\bar e}=e\; ,\num
$$
satisfying $a^*a=I_n$, $e^te=I_q$, and $\det e =1$. $\quad\square$

\subsec
The type III superdomains admit an action of the Lie supergroup
$SO^*(2n|q)$, which is defined as the intersection of $SU(n,n|q)$ with
the orthosymplectic supergroup $OSp(n|q)$. The latter is defined again
in terms of supermatrices of the form \fullgamma , where the submatrices
have the
same dimensions as in the case of $Sp(n|q)$. We require that
$\ber (\gamma)=1$, and
$$
\gamma^T L\gamma=L\; ,\ref{\osp}
$$
where $L$ is the supermatrix
$$
L=\left(\matrix{0 & I_n & 0\cr I_n & 0 & 0\cr 0 & 0 & \tau_q\cr}\right)\; ,
\num
$$
with $\tau_q$ defined in \taudef\ . Note that $L=L\star = L\inv$.
Solving the relations \gjgeq\
and \osp\ we write the generators of $SO^*(2n|q)$ in the form
$$
\gamma=\bordermatrix{&\sml n&\sml n&\sml q\cr
\sml n& a & b & \rho \cr \sml n& -{\bar b} & {\bar a} &
{\bar\rho}\tau \cr \sml q& \alpha & -\tau {\bar\alpha} & e \cr}\; ,
\qquad {\bar e}=\tau e\tau\; ,\num
$$
with the entries satisfying
$$
\eqalign{
& a^t{\bar b}+b^* a+\alpha^t\tau\alpha=0\; ,\cr
& a^t{\bar a}-b^*b+\alpha^t{\bar\alpha}=I_n\; ,\cr
& a^t{\bar\rho}\tau-b^*\rho-\alpha^t\tau e=0\; ,\cr
& \rho^t {\bar\rho}\tau-\tau\rho^* \rho + e^t\tau e=\tau\;
.\cr}\num
$$

We now consider the morphism $\smooth{\scdiii}\to
\smooth{SO^*(2n|q)}\widehat\otimes_\pi
\smooth{\scdiii}$ defined by \gactionii , where
$$
A:=a,\quad B:=(b,\rho),\quad C:=\left(\matrix{-{\bar b}\cr \alpha}\right),
\quad D:=\left(\matrix{{\bar a} & {\bar\rho}\tau\cr
-\tau {\bar\alpha} & e\cr}\right)\; .\num
$$

\prop\actiii{The above mophism defines a transitive action of
$SO^*(2n|q)$ on $\scdiii$.  Furthermore,
$$
\scdiii \cong  SO^*(2n|q)\big/ U(m)\times Sp(q/2).  \ref{\superquoiii}
$$}

\noindent{\it Proof.}
The proof parallels the proof of \actii. We write
$$
(I_n,Z') = (ZB\star + A\star)\inv (I_n, Z) \gamma\star.  \num
$$
The defining condition of $\scdiii$ is
$$
(I_n, Z)L^T \pmatrix {I_n\cr Z^T} = 0, \num
$$
which is preserved because $\gamma\star L^T (\gamma\star)^T = L^T$.
To prove \superquoiii, we note that the isotropy supergroup of $0$
consists of supermatrices
$$
\gamma=\left(\matrix{a & 0 & 0 \cr 0 & {\bar a} & 0 \cr
0 & 0 & e \cr}\right)\; ,\qquad {\bar e}=\tau e\tau\; ,\num
$$
satisfying $a^*a=I_n$, $e^t\tau e=\tau$, and $\det e=1$. $\quad\square$

\section\tdpssec{Triple determinants and Poisson structures}

\newsubsec
The construction of [6] rested on the framework of Jordan hermitian
triple systems.  For the purposes of this paper, we extract from this
framework the fact that the Bergman kernel of a Cartan domain is
given by
$$
K(z,w) = \Delta(z,w)^{-p},  \num
$$
where $\Delta(z,w)$ is a polynomial in $z$ and $\bar w$ (called the Jordan
triple determinant), and where $p$ is a positive integer called the genus of
the Cartan domain, see e.g. [6] (we plan to present the theory of Jordan
triples for Cartan superdomains elsewhere).  We let $\aut$
denote the Lie supergroup of superholomorphic automorphisms of $\cal D$.
The circular symmetry is a transformation of the form
$$
(z,\theta) \to (e^{i\varphi}z, e^{i\varphi/2}\theta), \num
$$
where $\varphi$ is a real number.

\subsec
For the quantization of superdomains, the central object will be an
analog of the triple determinant mentioned above.
We define a total genus $p=p_0 - p_1$, where $p_0$
is the genus of the underlying ordinary domain and $p_1$ is a
non-negative integer which we call the fermionic
genus.  Also for $\gamma\in\aut$ we define
$$
\gamma'(Z)_{\mu\nu} =
{\del \over \del Z_\mu} \gamma(Z)_\nu.  \ref{\gammap}
$$
In this definition, and throughout this paper,
derivatives with respect to odd variables are left derivatives, i.e.
$$
{\del\over \del\theta_1}\; (\theta_1\theta_2) = \theta_2. \num
$$
For future reference we note here that the chain rule takes the
following forms:
$$
\eqalign{&{\del\over \del Z_\mu} f\circ\gamma(Z) =
\sum_\rho \gamma'(Z)_{\mu\rho} {\del f\over\del Z_\rho} (\gamma(Z)), \cr
&{\del\over \del \ol Z_\mu} f\circ\gamma(Z) =
\sum_\rho (-1)^{\epsilon_\mu(\epsilon_\rho+1)}
\ol{\gamma'(Z)_{\mu\rho}} {\del f\over\del\ol Z_\rho} (\gamma(Z)), \cr} \num
$$
where $\epsilon_\mu := p(Z_\mu)$.  The extra sign in the second relation
occurs because
$$
{\del\over \del \ol Z_\mu} \ol{\gamma(Z)_\rho} = (-1)^{\epsilon_\mu
(\epsilon_\rho+1)} \ol{\gamma'(Z)_{\mu\rho}}.  \num
$$

\medskip
\thm\stdthm{For a Cartan superdomain there exists a polynomial
$N(Z,W)$ in $Z$ and $\ol W$ such that for all $\gamma\in\aut$,
$$
N(\gamma(Z),\gamma(W))^p = \ber \gamma'(Z) \> N(Z,W)^p
\>\ol{\ber \gamma'(W)}.  \num
$$
Furthermore,
$$
N(Z,W) = 1 - \sum_\mu \beta_\mu\inv Z_\mu \ol W_\mu
+ \hbox{\it higher order terms}, \ref{\nonsingref}
$$
where $\beta_\mu\inv$ are positive integers.}

\medskip
The polynomial $N(Z,W)$ is the super analog of the Jordan triple determinant.
Note that $N(Z,W)$ is invariant under the circular symmetry.
The theorem below states that $N(Z,W)$ has a simple transformation property
under $\aut$, a fact which will play an important role in the following.

\thm\agthm{There exists a unique holomorphic polynomial $a_\gamma(Z)$
such that:
\item{(i)} The automorphy factor $\ber\gamma'(Z)$ is given by
$$
\ber\gamma'(Z) = \ag(Z)^p\>;  \num
$$
\item{(ii)} We have the cocycle condition
$$
a_{\gamma_1\gamma_2}(Z) = a_{\gamma_1}(\gamma_2(Z)) a_{\gamma_2}(Z)\>;
\ref{\agcomp}
$$
\item{(iii)} The polynomial $N(Z,W)$ tranforms according to
$$
N(\gamma(Z),\gamma(W)) = a_\gamma(Z) \> N(Z,W)
\>\ol{a_\gamma(W)}\>.  \ref{\ntrans}
$$}

\medskip\noindent
We will prove \stdthm\ and \agthm\ in the next section.

For the following we define the Lebesgue measure $dz := d^{2n_0}z =
\prod_{l=1}^{n_0} {i\over 2} dz_l \wedge d\bar z_l$.  We also define
the Berezin integral
$d\theta := d^{n_1}\theta\> d^{n_1}\bar\theta$, which is normalized so that
$$
\int \prod_{l=1}^{n_1} (\bar\theta_l\theta_l)\; d\theta =1. \ref{\bercon}
$$
Let $dZ := dz\>d\theta$.  The Berezinian was defined precisely so that
if $Z' = \gamma(Z)$, then
$$
dZ' = \ber \gamma'(Z) dZ.  \num
$$

\cor\invthm{The measure
$$
d\mu(Z) :=  N(Z,Z)^{-p} dZ \ref{\invmdef}
$$
is invariant under the action of $\aut$.}

\subsec
The superalgebra $\smooth\scd$ of smooth functions on a Cartan superdomain
can be equipped with an $\aut$-invariant super Poisson structure.
This arises as follows. Let $\Omega^{k,l}(\scd)$, $k,l\in\bZ$, denote the
$\smooth\scd$-modules of forms of type $(k,l)$ on $\scd$, and let
$$
\del:\> \Omega^{k,l}(\scd) \to \Omega^{k+1,l}(\scd), \num
$$
and
$$
\dbr:\> \Omega^{k,l}(\scd) \to \Omega^{k,l+1}(\scd), \num
$$
denote the natural generalizations of the usual $\del$ and $\dbr$
operators.
We consider the even two-form defined by
$$
\eqalign{\omega(Z)&:= \del\dbr \log N(Z,Z) \cr
&= \sum_{\mu,\nu}\; (-1)^{\epsilon_\mu+1} d{\ol Z}_\mu\wedge
dZ_\nu
\;{\del^2\over \del Z_\nu \del\ol Z_\mu}
\log N(Z,Z) , \cr} \ref{\omegadef}
$$
where $\epsilon_\mu := p(Z_\mu)$.  The parity conventions for forms
and vector fields are $p(dZ_\mu) = \epsilon_\mu + 1$,
$p(\del/\del Z_\mu) = \epsilon_\mu$.

\prop\omprop{$\omega$ is an $\aut$-invariant supersymplectic form on
$\scd$.}

\medskip\noindent{\it Proof.}
To see that $\omega$ is $\aut$-invariant, we note that, as a consequence of
\agthm,
$$
\log N(\gamma(Z),\gamma(Z))   =  \log N(Z,Z)  +
\log\ag(Z) + \log{\ol {\ag(Z)}}.  \num
$$
Since $a_\gamma(Z)$ is holomorphic,
$$
\del\dbr \log\ag(Z)  = \del\dbr\log{\ol {\ag(Z)}} = 0,  \num
$$
and so $\gamma^\ast\omega = \omega$, as claimed.

Since $d = \del + \dbr$, it follows immediately that $d\omega = 0$.
It remains to show that $\omega$ is non-degenerate. Owing to the
$\aut$-invariance, it is sufficient to prove that $\omega(0)$ is
non-degenerate. This, however, is clear since \nonsingref\ implies
that
$$
\omega(0)=\sum_{\mu} \beta_\mu\inv d\ol Z_\mu\wedge dZ_\mu\; .
\qquad\square
$$

\bigskip
In components we write the symplectic form as
$$
\omega_{\mu\nu}(Z) = (-1)^{\epsilon_\mu}
{\del^2\over \del Z_\nu \del\ol Z_\mu} \log N(Z,Z)\inv, \num
$$
so that $\omega(Z) = \sum_{\mu,\nu} d\ol Z_\mu \wedge dZ_\nu
\> \omega_{\mu\nu}(Z)$.

\bigskip
We now construct the super Poisson bracket associated to $\omega$.
The Poisson bracket is defined by the inverse of $\omega$
with respect to the natural pairing
$$
\Omega^{1,1} \otimes \Omega^{-1,-1} \to \bC, \num
$$
which sends
$$
d\ol Z_\mu \wedge dZ_\nu \otimes {\del\over \del\ol Z_\sigma}
\wedge {\del\over \del Z_\rho} \mapsto \delta_{\nu\sigma}
\delta_{\mu\rho}.  \num
$$
We require $\omega\otimes \omega\inv \mapsto 1$.  Note that
this corresponds to $\sum_{\nu} \omega_{\mu\nu}(Z)
\omega^{-1}_{\phantom{-1}\nu\rho}(Z) = \delta_{\mu\rho}$.
Then the Poisson bracket is defined by
$$
\{f,g\} := \omega\inv(Z) (df,dg).  \num
$$
According to Theorems 5.4 and 5.5 of [3], the bracket $\{\cdot,\cdot\}$
defined in this way indeed has the properties of a super Poisson
bracket, as formulated in the Introduction.

Using the invariance of $\omega$, we can write the Poisson
bracket more conveniently.
To each $Z\in\scd$ we associate an element $\gz\in
{\rm Aut}(\scd)$ such that $\gz(0) = Z$.
Let $\pi\in \Omega^{-1,-1}(\scd)$ be defined by
$$
\pi(Z):= \sum_{\mu,\nu} P(Z)_{\mu\nu}\;
{\partial\over{\partial Z_\nu}}\wedge{\partial\over{\partial\ol Z_\mu}}\;,
\ref{\bivectref}
$$
where
$$
P(Z)_{\mu\nu} :=  \sum_\rho
\beta_\rho \ol{\gz'(0)_{\rho\mu}} \gz'(0)_{\rho\nu} .  \ref{\bigpdef}
$$

\thm\spalg{The Poisson bracket associated to $\omega$ is given by
$$
\{f,g\}=\pi(df,dg).\ref{\poissonref}
$$
Consequently, the pair $(C^{\infty}(\scd),\>\{.,.\})$ is a
Poisson superalgebra with an ${\rm Aut}(\scd)$-invariant Poisson bracket.}

\medskip\noindent
{\it Proof.} We make use of the invariance property by inverting $\omega$
at the origin and then pushing forward by the action of the supergroup
$\aut$.  Clearly,
$$
\omega^{-1}(0)=\sum_{\mu} \beta_\mu {\partial\over{\partial Z_\mu}}
\wedge{\partial\over{\partial\ol Z_\mu}}\;,\num
$$
and so
$$
\{f,g\}(0) = \sum_\mu (-1)^{\epsilon_\mu p(f)} \beta_\mu
\left[ {\del f\over \del
Z_\mu}(0){\del g\over \del \ol Z_\mu}(0) - (-1)^{\epsilon_{\mu}}
{\del f\over \del \ol Z_\mu}(0){\del g\over \del Z_\mu}(0) \right],
\ref{\pbatzeroref}
$$
where $\epsilon_\mu := p(Z_\mu)$. From the invariance of $\omega$
under
${\rm Aut}(\scd)$ we conclude that
$$
\eqalign{
\{f,g\}(Z) := \omega^{-1}(Z)(df,dg) &= \omega^{-1}(\gz(0))
\bigl(d(f\circ\gz),d(g\circ\gz)\bigr)\cr
&=\omega^{-1}(0) \bigl(d(f\circ\gz),d(g\circ\gz)\bigr)\cr
&=\{f\circ\gz, g\circ \gz\}(0).\cr}
$$
Consequently, using \pbatzeroref\ we obtain that
$$
\eqalign{
\{f,g\}(Z)&=\sum_{\rho,\mu,\nu}  \beta_\rho
\ol{\gz'(0)_{\rho\mu}} \gz'(0)_{\rho\nu} \cr
&\quad\times(-1)^{\epsilon_\nu p(f)} \left[
{\del f\over \del Z_\nu}(Z) {\del g\over \del \ol Z_\nu}(Z) -
(-1)^{\epsilon_\mu \epsilon_\nu} {\del f\over \del \ol Z_\mu}(Z)
{\del g\over \del Z_\nu}(Z) \right] .\cr} \ref{\spbdef}
$$
In view of \bigpdef\ we obtain
$$
\eqalign{
&\{f,g\}(Z)\cr
&\quad =\sum_{\mu,\nu} P(Z)_{\mu\nu} (-1)^{\epsilon_\mu p(f)}
\left[  {\del f\over \del Z_\nu}(Z)
{\del g\over \del \ol Z_\mu}(Z) -
(-1)^{\epsilon_\mu\epsilon_\nu} {\del f\over \del \ol Z_\mu}(Z)
{\del g\over \del Z_\nu}(Z) \right]\cr
&\quad =\pi(Z)(df,dg)\; ,\cr}
$$
as claimed.  $\quad\square$

\cor\omegacor{The inverse of $\omega$ is given by
$$
\omega^{-1}_{\phantom{-1}\mu\nu}(Z) = P_{\mu\nu}(Z), \num
$$
and as a consequence
$$
\ber \omega(Z) =  N(Z,Z)^{-p}\; \prod_\mu \beta_\mu\inv , \num
$$
where $\omega_{\mu\nu}$ is viewed as a supermatrix.}

\medskip\noindent{\it Proof.}
The first statement is the content of the previous theorem.
The definition of $P_{\mu\nu}$ then implies that
$$
\ber\omega(Z) =  |\ber \gz'(0)|^{-2}\;\prod_\mu \beta_\mu\inv .  \num
$$
Applying \stdthm\ to $\gz$ yields
$$
N(Z,Z)^p = N(\gz(0),\gz(0))^p = |\ber \gz'(0)|^2,  \num
$$
and the second statement follows. $\quad\square$

\subsec
For $\sigma\in \Omega^{-1,-1}(\scd)$ given by
$$
\sigma = \sum_{\mu,\nu} f_{\mu\nu}(Z) {\del\over \del Z_\nu} \wedge
{\del\over \del \ol Z_\mu},  \num
$$
the map $\del : \Omega^{-1,-1}(\scd) \to
\Omega^{0,-1}(\scd)$ takes $\sigma$ to
$$
\del\sigma = \sum_{\mu,\nu} (-1)^{\epsilon_\nu(\epsilon_\mu+1)}
{\del f_{\mu\nu} \over \del Z_\nu}(Z) \;{\del\over \del \ol Z_\mu}.
\ref{\dsexp}
$$

\thm\dsthm{The two-vector field $\sigma\in \Omega^{-1,-1}$ defined by
$$
\sigma = \sum_{\mu,\nu}\; {P_{\mu\nu}(Z)\over N(Z,Z)^p}\;
{\del\over \del Z_\nu} \wedge
{\del\over \del \ol Z_\mu},  \num
$$
satisfies $\del\sigma = 0$.}

\medskip\noindent{\it Proof.}
For convenience in this proof let $\del_\mu := {\del\over \del Z_\mu}$
and likewise for $\dbr_\mu$.
We start with the fact that $P_{\mu\nu} =
\omega\inv_{\phantom{-1}\mu\nu}$, so that
$$
\del_\rho P_{\mu\nu} = - \sum_{\alpha,\beta} (-1)^{\epsilon_\rho
(\epsilon_\mu + \epsilon_\alpha)}  P_{\mu\alpha}
(\del_\rho \omega_{\alpha\beta}) P_{\beta\nu}.  \num
$$
Thus
$$
\sum_\nu (-1)^{\epsilon_\nu(\epsilon_\mu+1)} \del_\nu P_{\mu\nu}
= - \sum_{\nu,\alpha,\beta} (-1)^{\epsilon_\nu(\epsilon_\alpha+1)}
P_{\mu\alpha} (\del_\nu \omega_{\alpha\beta}) P_{\beta\nu}.  \num
$$
Now the statement that $\del\omega = 0$ means that
$\del_\nu\omega_{\alpha\beta} = (-1)^{\epsilon_\nu \epsilon_\alpha}
\del_\alpha\omega_{\nu\beta}$, so that
$$
\sum_\nu (-1)^{\epsilon_\nu(\epsilon_\mu+1)} \del_\nu P_{\mu\nu}
= - \sum_{\nu,\alpha,\beta} (-1)^{\epsilon_\nu}
P_{\mu\alpha} (\del_\alpha \omega_{\nu\beta}) P_{\beta\nu}.
\ref{\sumnu}
$$
By the definitions of the supertrace and the Berezinian [3],
$$
\eqalign{
\sum_{\nu,\beta} (-1)^{\epsilon_\nu} (\del_\alpha \omega_{\nu\beta})
\omega\inv_{\phantom{-1}\beta\nu} &= \del_\alpha \mathop{\rm Str} \log \omega
\cr
&= \del_\alpha \log \ber \omega. \cr}  \num
$$
By \omegacor\ we see that $\ber \omega$ is equal a constant
times $N(Z,Z)^{-p}$.  Thus
$$
\del_\alpha \log \ber \omega = -p \> \del_\alpha \log N.  \num
$$
Returning to \sumnu, we have
$$
\sum_\nu (-1)^{\epsilon_\nu(\epsilon_\mu+1)} \del_\nu P_{\mu\nu}
= p \sum_\alpha P_{\mu\alpha} \del_\alpha \log N.  \num
$$

In view of the explicit formula \dsexp, the statement that
$\del\sigma = 0$ is equivalent to
$$
\sum_\nu (-1)^{\epsilon_\nu(\epsilon_\mu+1)} \del_\nu
{P_{\mu\nu}\over N^p} = 0, \num
$$
for all $\mu$.  Using the results of the last paragraph we evaluate
$$
\eqalign{\sum_\nu (-1)^{\epsilon_\nu(\epsilon_\mu+1)} \del_\nu
{P_{\mu\nu}\over N^p} &= pN^{-p} \sum_\alpha P_{\mu\alpha}
\del_\alpha\log N + \sum_\nu P_{\mu\nu} \del_\nu N^{-p} \cr
&= 0. \qquad\square\cr} \num
$$

\section\proofsec{Proof of Theorems V.1 and V.2}

\newsubsec
In this section we define the ``super triple determinant'' $N(Z,W)$ for
matrix superdomains and establish Theorems V.1 and V.2.
We will prove these theorems after establishing a series
of propositions.

\lemma\azbczd{For $\gamma\in\sgpi$,
$$
\det (A\star + ZB\star) = \ber (CZ+D), \num
$$
where $A,B,C$, and $D$ are the matrix blocks of $\gamma$.}

\medskip\noindent{\it Proof.}
Using \iwzizw\ we have
$$
\eqalign{\ber (CZ+D) &= \ber D \; \ber(\inq + D\inv CZ)\cr
&= \ber D \; \det(I_m + ZD\inv C).\cr} \num
$$
Using \grelns\ we obtain
$$
\eqalign{I_m + ZD\inv C &=
A\star A - C\star C + ZD\star C + ZB\star B D\inv C\cr
&= A\star A - A\star BD\inv C + ZB\star A + ZB\star BD\inv C\cr
&= (A\star + ZB\star)(A - BD\inv C).\cr} \num
$$
We now combine this with the fact that
$$
\ber \gamma = \det(A - BD\inv C) \ber D = 1  \num
$$
to see that
$$
\ber (CZ+D) = \det (A\star + ZB\star).  \qquad\square\num
$$

\prop\gpzpropi{For $\gamma\in SU(m,n|q)$ acting on $\scdi$,
$$
\ber \gamma'(Z) = {1\over\det (A\star + ZB\star)^{m+n-q}}.  \num
$$}
\medskip\noindent{\it Proof.}
The matrix of derivatives can be evaluated explicitly,
$$
{\del Z'_{ij}\over\del Z_{mn}} = (ZB\star + A\star)\inv_{im}
( D\star - B\star Z' )_{nj} . \num
$$
In the matrix notation of \gammap\ we write
$$
\gamma'(Z) = [(ZB\star + A\star)^{-1}]^T \otimes (D\star - B\star Z').
\num
$$
Using the relations \grelns, we see that
$$
\eqalign{D\star - B\star Z' &= D\star - B\star (AZ + B) (CZ+D)\inv\cr
&= \bigl[ D\star CZ + D\star D - B\star AZ - B\star B\bigr]
(CZ+D)\inv\cr
&= (CZ+D)\inv. \cr} \num
$$
Thus the matrix of derivatives becomes
$$
\gamma'(Z) = [(ZB\star + A\star)^{-1}]^T \otimes
(CZ+D)\inv , \num
$$
and its Berezinian is
$$
\ber \gamma'(Z) = \det (ZB\star + A\star)^{-(n-q)}
\ber (CZ+D)^{-m} . \num
$$
The proposition follows from \azbczd. $\quad\square$

\medskip
\prop\gpzpropii{For $\gamma\in Sp(n|q)$ acting on $\scdii$,
$$
\ber\gamma'(Z) = {1\over\det (A\star + ZB\star)^{n+1-q}}\;.  \num
$$}

\medskip\noindent{\it Proof.}
First we study the case when $Z=0$.
Choosing coordinates $Z_{ij}$, where either $1\le i<j\le n$ or
$j>n$, the supermatrix $\gamma'(0)$ is given by
$$
\eqalign{
&\gamma'(0)_{kl,ij} = {\del Z'_{ij}\over \del Z_{kl}} \cr
&= \bordermatrix{&\sml{i\le j \le n}&\sml{j>n}\cr
\sml{k\le l \le n}&{1\over 1 + \delta_{kl}}
\bigl[(A^{\ast\inv})_{ik} u_{lj} +
(A^{\ast\inv})_{il} u_{kj}\bigr] &
{1\over 1+ \delta_{kl}} \bigl[(A^{\ast\inv})_{ik} \sigma_{lj} +
(A^{\ast\inv})_{il} \sigma_{kj}\bigr]\cr
\sml{l>n}&-(A^{\ast\inv})_{ik} \eta_{lj}&
(A^{\ast\inv})_{ik} v_{lj}\cr}, \cr}  \num
$$
where we have represented the block entries of $D\inv$ by
$$
D\inv = \pmatrix{u&\sigma\cr \eta& v\cr}.  \num
$$
Writing $\gamma'(0)$ as
$$
\gamma'(0) = \pmatrix{T_1&T_2\cr T_3&T_4\cr},  \num
$$
we need to compute
$$
\ber T = \det (T_1 - T_2T_4\inv T_3) \det T_4\inv.  \ref{\needco}
$$
We start by observing that
$$
[T_4\inv T_3]_{kl,ij} = \delta_{ik} [v\inv\eta]_{lj},  \num
$$
so that
$$
\eqalign{\bigl[T_1 - T_2T_4\inv T_3\bigr]_{kl,ij} &= {1\over
1+ \delta_{kl}} \Bigl[ (A^{\ast\inv})_{ik}
(u - \sigma v\inv \eta)_{lj} +
(A^{\ast\inv})_{il} (u - \sigma v\inv \eta)_{kj}
\Bigr]  \cr
&\equiv \bigl[\bar A\inv \otimes_s
(u - \sigma v\inv \eta) \bigr]_{kl,ij}, \cr} \ref{\symtens}
$$
where $A\otimes_s B$ denotes the symmetric tensor product of the
matrices $A$ and $B$. Now from \nstdf\ we see that
$$
(u - \sigma v\inv \eta) = \bar A\inv,  \num
$$
so we have
$$
\det (T_1 - T_2T_4\inv T_3)  = {1\over \det (\bar A\otimes_s \bar A)}
= (\det A\star)^{-(n+1)}.  \num
$$
To complete the calculation of \needco\ we have
$$
\det T_4 = \det (A\star)^{-q} \det v^{-n}.  \num
$$
In terms of $D$, $v = (e + \bar\alpha \bar a\inv \bar\rho)\inv$, and one
easily sees from the relations that $v^tv = I_q$.
The result is thus
$$
\ber \gamma'(0) = (\det A\star)^{-(n+1 - q)}.  \ref{\gpzero}
$$

To complete the proof we consider the case where $\gamma$ maps $Z$ to
$Z'\ne 0$.  Let $\gamma = \gamma_2 \circ \gamma_1$, where
$$
\gamma_1(Z) = 0, \qquad \gamma_2(0) = Z',  \ref{\gzgzp}
$$
We write
$$
\gamma_i = \pmatrix{A_i& B_i\cr C_i& D_i\cr},\quad i=1,2, \num
$$
and
$$
\gamma = \pmatrix{A&B\cr C&D\cr} = \pmatrix{A_2A_1 + B_2C_1&
A_2B_1 + B_2D_1\cr C_2A_1 + D_2C_1 & C_2B_1 + D_2D_1\cr}. \num
$$
Because of \gzgzp,
$$
ZD_1\star + C_1\star = 0, \num
$$
so that
$$
\eqalign{A\star + ZB\star &= A_1\star A_2\star  + C_1\star B_2\star
+ ZB_1\star A_2\star + ZD_1\star B_2\star \cr
&= (A_1\star + ZB_1\star) A_2\star \cr
&= (A_1\star + C_1\star (D_1\star)\inv B_1\star) A_2\star. \cr} \num
$$
Applying the result \gpzero\ and the fact that $(A_1- B_1D_1\inv
C_1)\inv$
is the upper right submatrix of $\gamma_1\inv$, we have
$$
\det (A\star+ZB\star)^{-p} = {\ber\gamma_2'(0) \over
\ber(\gamma_1\inv)'(0)}
= \ber \gamma_2'(0) \ber \gamma_1'(Z) = \ber \gamma'(Z).
\qquad\square\num
$$

\medskip
\prop\gpzpropiii{For $\gamma\in SO\star(2n|q)$ acting on $\scdiii$,
$$
\ber \gamma'(Z) = {1\over\det (A\star + ZB\star)^{n-1-q}}.  \num
$$}

\medskip\noindent{\it Proof.}
The proof follows closely that of \gpzpropii.  In place of equation
\symtens, we obtain
$$
\eqalign{\bigl[T_1 - T_2T_4\inv T_3\bigr]_{kl, ij} &= {1\over
1+ \delta_{kl}} \Bigl[ (A^{\ast\inv})_{ik}
(u - \sigma v\inv \eta)_{lj} -
(A^{\ast\inv})_{il} (u - \sigma v\inv \eta)_{kj}
\Bigr]  \cr
&\equiv \bigl[\bar A\inv \otimes_a
(u - \sigma v\inv \eta) \bigr]_{kl,ij}, \cr} \ref{\symtens}
$$
where $A\otimes_a B$ denotes the antisymmetric tensor product of the
matrices $A$ and $B$. Since $\det A\star\otimes_a A\star =
(\det A\star)^{n-1}$, we thus obtain
$$
\gamma'(0) = (\det A\star)^{-(n-1)+q},  \num
$$
in place of \gpzero.  The second half of the proof is then identical to
that above. $\quad\square$

\bigskip
Based on the preceeding four propositions, for all three types we define
the super triple determinant
$$
N(Z,W):=\det(I_m - Z\star W) = \ber {(I_{n|q}-W^*Z)} ,\num
$$
and the transformation factor
$$
a_\gamma(Z):=\det (A\star + ZB\star) = \ber {(CZ+D)} .\ref{\agdef}
$$

\medskip
\prop\gtransprop{With the above definitions,
$$
N( \gamma(Z),\gamma(W)) = \ag(Z) N(Z,W) \ol{\ag(W)}.  \num
$$}
{\it Proof.}
The statement is that
$$
\ber \bigl(\inq - \gamma(W)\star \gamma(Z)\bigr) = \ag(Z) \ber (\inq -
W\star Z) \ol{\ag(W)}.  \num
$$
The defining property \gjgeq\ of $\sgpi$ implies that
$$
\inq - \gamma(W)\star \gamma(Z) = {(CW+D)\star}\inv \bigl(\inq -
W\star Z\bigr) (CZ+D)\inv.  \num
$$
The proposition then follows from \agdef.  $\quad\square$

\subsec
{\it Proof of Theorems V.1 and V.2.}
Theorem V.2 (i) is established in \gpzpropi, \gpzpropii, and
\gpzpropiii\ for types I, II, and III, respectively (incidentally,
the fermionic genus $p_1$ turns out to be equal
to $q$ in all these cases). Part (iii) of Theorem V.2 and
the first statement of Theorem V.1 are proven in
\gtransprop.  The second statement of Theorem V.1 is clear.
In particular, we find that $\beta^{-1}_\mu =1$ or $2$ in \nonsingref.

It remains to prove property (ii) of Theorem V.2.  Let
$$
\gamma_i = \pmatrix{A_i & B_i\cr C_i& D_i\cr}, \num
$$
for $i = 1,2$.  We have
$$
\eqalign{
a_{\gamma_1\gamma_2}(Z) &= \ber \Bigl[(C_1A_2 + D_1C_2)Z +
(C_1B_2
+ D_1D_2) \Bigr]\cr
&= \ber \Bigl[ C_1(A_2Z + B_2) + D_1(C_2Z + D_2)\Bigr] \cr
&= \ber (C_1\gamma_2(Z) + D_1) \ber (C_2Z + D_2) \cr
&= a_{\gamma_1}(\gamma_2(Z)) a_{\gamma_2}(Z). \quad\square \cr}
\num
$$

\subsec
For future reference, we give here explicit formulas for the group elements
$\gz$.  For type I, $\gz$ can be written as
$$
\gz := \left(\matrix{I_m&Z\cr Z\star & \inq}\right) \left(\matrix{A&0\cr
0& D}\right),  \ref{\gzdefi}
$$
for any $A$ and $D$ which satisfy
$$
\eqalign{&AA\star = (I_m - ZZ\star)\inv,  \cr
&DD\star = (\inq - Z\star Z)\inv,  \cr} \ref{\adsat}
$$
and $\ber D\star \>\det A = 1$.

For type II we have
$$
\gz := \left(\matrix{I_n&z^t&\sigma \cr
z\star& I_n & -\bar\sigma\cr
\theta\star&\theta^t& I_q\cr}\right)
\left(\matrix{a&0&0\cr 0&\bar a&0\cr 0&0& e\cr} \right),  \num
$$
where
$$
\eqalign{aa\star &= (I_n - ZZ\star)\inv, \cr
\sigma &= (I_n - z \bar z)\inv ( \theta-z \bar\theta), \cr
ee^t &= (I_q + \sigma^t \bar\sigma - \sigma\star\sigma)\inv ,\cr} \num
$$
and where
$$
\det e = {\det (I_n - ZZ\star)\over \det (I_n - z\bar z)} .\num
$$

Finally, for type III,
$$
\gz := \left(\matrix{I_n&- z^t&\sigma \cr
z\star& I_n & \bar\sigma\tau\cr
\theta\star& -\tau\theta^t& I_q\cr}\right)
\left(\matrix{a&0&0\cr 0&\bar a&0\cr 0&0& e\cr} \right),  \num
$$
where
$$
\eqalign{aa\star &= (I_n - zz\star - \theta\theta\star)\inv, \cr
\sigma &= (I_n + z \bar z)\inv (\theta+ z\bar\theta\tau),  \cr
ee\star &= (I_q + \tau\sigma^t \bar\sigma\tau -
\sigma\star\sigma)\inv ,\cr} \num
$$
and where
$$
\det e = {\det (I_n - ZZ\star)\over \det (I_n + z\bar z)} .\num
$$

\section\qntsec{Quantization}

\newsubsec
Our framework for the quantization of a Cartan superdomain $\scd$ rests
on the following perturbation of the invariant measure. We will show later
that there is $r_0(\scd)>0$ such that the measure $N(Z,Z)^r d\mu(Z)$
has a finite volume for $r\geq r_0(\scd)$.
We set
$$
\dmu Z := \Lambda_r N(Z,Z)^r d\mu(Z)=\Lambda_r N(Z,Z)^{r-p} dz\; d\theta ,
\ref{\dmudef}
$$
for $r\ge r_0(\scd)$, where $d\mu$ is the invariant measure of \invthm\ and
$\Lambda_r$ is chosen so that the total integral is normalized to one.

For $f$ and $g\in B^{\infty}(\scd)$, we set
$$
(f,g)_r := \int_\scd \ol{f(Z)} g(Z) \dmu Z.  \num
$$
This form is not positive definite and so it does not define an inner
product on $B^{\infty}(\scd)$. The crucial property of $(\cdot,\cdot)_r$
is, however, that its restriction to the subspace of superholomorphic
functions is positive definite. In fact, a more general
property holds (which we will need). We consider the superspace
$B^{\infty}_\ast(\cal D)$
of functions $f$ for which $\partial f/\partial{\bar\theta_j}\; =0$.
Observe that this notion is not invariant under superholomorphic changes
of coordinates on $\scd$. The following theorem will be proven in
the next section.

\medskip
\thm\posipthm{There exists $r_0(\scd)>0$ such that for all
$r\geq r_0(\scd)$, the sesquilinear form $(\cdot,\cdot)_r$ defines
an inner product on $B^{\infty}_\ast(\cal D)$.}

\medskip
Consider the space ${\rm Hol}(\scd)$ of superholomorphic functions in
$B^{\infty}(\scd)$. As a consequence of the above theorem, $(\cdot,\cdot)_r$
is an inner product on this space. The completion of ${\rm Hol}(\scd)$ in
the norm induced by this inner product forms a Hilbert space, which we
denote by $\hil$.

\subsec
In this subsection we state some facts concerning the measure
$d\mu_r$ that will be useful later.

\prop\dmuprop{The form \dmudef\ has the transformation property
$$
\dmu {\gamma(Z)} = \left[\ag(Z) \ol{\ag(Z)} \right]^r
\dmu Z,  \num
$$
for and $\gamma\in\aut$.}

\medskip\noindent{\it Proof.}
This is a direct consequence of \agthm. $\quad\square$

\prop\fintprop{There is a constant $C>0$ such that for
$r$ sufficiently large
$$
\int N(Z,Z)^{r-p} d\theta
= C r^{n_1} \Delta(z,z)^{r-p_0} [1 + O(r\inv)],  \num
$$
uniformly in $z$, where $\Delta(z,z)$ is the triple determinant of
the underlying domain.}

\medskip\noindent
\fintprop\ will be established in Section \possec.

\prop\normprop{The normalization constant $\Lambda_r$ has the
behavior
$$
\Lambda_r = C r^{n_0 - n_1} [1 + O(r\inv)],  \num
$$
as $r\to\infty$.}
\medskip\noindent{\it Proof.}
The statement follows immediately from \fintprop\ and Lemma 3.1 (i)
of [6]. $\quad\square$

\subsec
The Hilbert space $\hil$ carries a natural projective unitary representation
of $\aut$.  This is given by $\gamma \to U(\gamma)$,
where
$$
U(\gamma\inv)\phi(Z) = \ag(Z)^r
\phi(\gamma(Z)).  \ref{\ugdef}
$$
Clearly, each $U(\gamma\inv)$ is unitary because of \dmuprop.
We see that $U$ is a projective representation as follows.

For notational convenience in the following argument,
we write $a(\gamma, Z)$ in place of $a_\gamma(Z)$.
For $\gamma_1, \gamma_2 \in \aut$, define the function
$$
\lambda(\gamma_1, \gamma_2)(Z) := {1\over 2\pi i} \Bigl\{
\log a(\gamma_2\inv \gamma_1\inv, Z) - \log a(\gamma_1\inv, Z)
- \log a(\gamma_2\inv, \gamma_1\inv(Z)) \Bigr\}.  \ref{\lambdaref}
$$
\thm\cocyclethm{The function $\lambda(\gamma_1, \gamma_2)$ defined
above has the following properties:
\item{(i)} $\lambda(\gamma_1, \gamma_2)(Z)$ does not depend on $Z$.
Thus $\lambda(\gamma_1, \gamma_2)$ is a function on $\aut\times\aut$.
\item{(ii)} We have the following cocycle condition:
$$
\lambda(\gamma_1,\gamma_2\gamma_3) + \lambda(\gamma_2, \gamma_3)
- \lambda(\gamma_1\gamma_2, \gamma_3) - \lambda(\gamma_1, \gamma_2)
= 0.  \ref{\gsum}
$$
\item{(iii)} $\lambda(\gamma_1,\gamma_2)\in\{-1, 0, 1\}$.}

\medskip\noindent{\it Proof.}
(i) We take the gradient of $\lambda(\gamma_1, \gamma_2)(Z)$ as follows:
$$
\eqalign{2\pi i \nabla \lambda(\gamma_1, \gamma_2)(Z) &= {1\over
a(\gamma_2\inv\gamma_1\inv, Z)} \nabla a(\gamma_2\inv\gamma_1\inv, Z)
- {1\over a(\gamma_1\inv, Z)} \nabla a(\gamma_1\inv, Z)\cr
&\qquad- {1\over a(\gamma_2\inv, \gamma_1\inv(Z))} \nabla
a(\gamma_2\inv, \gamma_1\inv(Z)) . \cr} \num
$$
By \agcomp,
$$
a(\gamma_2\inv\gamma_1\inv, Z) =
a(\gamma_2\inv, \gamma_1\inv(Z)) a(\gamma_1\inv, Z).  \num
$$
We thus see that
$$
\eqalign{{1\over a(\gamma_2\inv\gamma_1\inv, Z)} \nabla
a(\gamma_2\inv\gamma_1\inv, Z) &=
- {1\over a(\gamma_1\inv, Z)} \nabla a(\gamma_1\inv, Z) \cr
&\qquad - {1\over a(\gamma_2\inv, \gamma_1\inv(Z))} \nabla
a(\gamma_2\inv, \gamma_1\inv(Z)) . \cr} \num
$$

(ii) Consider the first two terms in \gsum.  In view of (i),
we can evaluate
either $\lambda$ at any point.  We choose to evaluate
the first $\lambda$ at $Z$ and the second at $\gamma_1\inv(Z)$.
The sum of these terms is thus
$$
\eqalign{&\lambda(\gamma_1,\gamma_2\gamma_3)(Z) + \lambda(\gamma_2, \gamma_3)
(\gamma_1\inv(Z))\cr
&\qquad= {1\over 2\pi i}
\Bigl\{ \log a(\gamma_3\inv \gamma_2\inv\gamma_1\inv, Z) -
\log a(\gamma_1\inv, Z) \cr
&\qquad\qquad -
\log a(\gamma_2\inv, \gamma_1\inv(Z)) -
\log a(\gamma_3\inv,  \gamma_2\inv(\gamma_1\inv(Z)))\Bigr\} \cr}
\ref{\leval}
$$
By adding and subtracting ${1\over 2\pi i} \log a((\gamma_1\gamma_2)\inv
, Z)$, we see that \leval\ is in fact equal to
$$
\lambda(\gamma_1\gamma_2, \gamma_3) + \lambda(\gamma_1, \gamma_2).
\num
$$

To prove (iii), we set $Z=0$ in \lambdaref\ and use \agcomp .\qquad
$\square$

\cor\repcor{Formula \ugdef\ defines a projective unitary
representation of $\aut$ on $\hil$.}

\medskip\noindent{\it Proof.}
Set
$$
\sigma(\gamma_1,\gamma_2):=
\exp\Bigl\{ 2\pi i r \lambda(\gamma_1, \gamma_2) \Bigl\}
\ref\sigdef
$$
As a consequence of \agthm,
$$
U(\gamma_1\gamma_2) = \sigma(\gamma_1,\gamma_2)U(\gamma_1)U
(\gamma_2).\num
$$
The cocycle condition,
$$
\sigma(\gamma_2,\gamma_3)\sigma(\gamma_1\gamma_2,\gamma_3)
^{-1}
\sigma(\gamma_1,\gamma_2\gamma_3)\sigma(\gamma_1,\gamma_2)
^{-1}=1,\num
$$
follows from \cocyclethm\ (ii), which shows that \ugdef\ is
consistent with associativity. The unitarity
is a consequence of \dmuprop. $\quad\square$

\subsec
A fundamental component of our construction is the Bergman (or
reproducing) kernel for the space $\hil$.  Let
$$
\rker ZW := N(Z,W)^{-r}.  \ref{\krdef}
$$

\prop\bergprop{The kernel function  \krdef\ has the reproducing property,
i.e. for $\phi\in\hil$,
$$
\phi(Z) = \int_\scd \rker ZW \phi(W) \dmu W.  \ref{\phkph}
$$}
\medskip\noindent
{\it Proof.}
We compute the right-hand side of \phkph\ by making the substitution $W =
\gz(Y)$, where $\gz$ is an element of Aut$(\scd)$ such that
$\gz(0) = Z$.  This yields
$$
\eqalign{&\int_\scd \rker ZW \phi(W) \dmu W \cr
&\qquad= \int_\scd \rker
Z{\gz(Y)} \phi(\gz(Y)) \dmu {\gz(Y)} \cr
&\qquad= \int_\scd \left[a_\gz(0) \ol{a_\gz(Y)}\right]^{-r}
\phi(\gz(Y))
\left[a_\gz(Y) \ol{a_\gz(Y)}\right]^r \dmu Y \cr
&\qquad= \int_\scd a_\gz(0)^{-r} a_\gz(Y)^r
\phi(\gz(Y)) \dmu Y.  \cr}  \num
$$
We apply the simple fact that for $\psi$ holomorphic,
$$
\int_\scd \psi(W) d\mu_r(W)= \psi(0),  \num
$$
which is a consequence of circular symmetry, and obtain
$$
\int_\scd a_\gz(0)^{-r} a_\gz(Y)^r
\phi(\gz(Y)) \dmu Y = \phi(Z).  \qquad\square\num
$$

\bigskip
For $g\in B^\infty(\scd)$, we define the projection $P$ by
$$
Pg(Z) := \int_\scd \rker ZW g(W) \dmu W.  \num
$$
Clearly, $Pg\in\hil$, and $Pg = g$ for $g\in\hil$.

\prop\kzwprop{
$$
\int_\scd \ol Z_\mu Z_\rho \dmu Z = {\beta_\mu\over r}\delta_{\mu\rho},
\ref{\murhoint}
$$
where $\beta_\mu$ is the constant of \stdthm.}

\medskip\noindent{\it Proof.}
Let
$$
A_{\mu\rho} = \int_\scd \ol Z_\mu Z_\rho \dmu Z. \num
$$
Because of \posipthm\ the matrix $A_{\mu\rho}$ is invertible.
Using \bergprop\ we can also write
$$
\int_\scd \ol Z_\mu Z_\rho \dmu Z =
\int_{\scd\times\scd} \ol{Z_\mu} \rker ZW W_\rho \dmu Z \dmu W. \ref{\zkwint}
$$
The circular symmetry and the expansion of $N(Z,W)$ in \stdthm\ imply that
the right-hand side of \zkwint\ is given by
$$
\sum_\nu {r\over\beta_\nu}  \int_{\scd\times\scd} \ol{Z_\mu}
Z_\nu \ol{W_\nu}  W_\rho \dmu Z \dmu W.  \num
$$
In terms of $A$ this implies
$$
A_{\mu\rho} = \sum_\nu A_{\mu\nu} {r\over\beta_\nu} A_{\nu\rho}.  \num
$$
We apply $A\inv$ to both sides of
this equation and obtain \murhoint. $\quad\square$

\subsec
As described in Section \stoesec\ we define super Toeplitz
operators $T_r(f)$ on $\hil$, for $f\in B^\infty(\scd)$, by
setting
$$
T_r(f)\phi(Z) := \int_\scd \rker ZW f(W) \phi(W) \dmu W.  \num
$$
The map $f\mapsto T_r(f)$ will be the quantization map in our scheme.
We first establish some basic properties of the super Toeplitz operators.

First of all, observe that
$$
T_r(f\circ \gamma) = U(\gamma)\inv T_r(f) U(\gamma),  \ref{\trfcg}
$$
where $U(\gamma)$ is defined by \ugdef.

Secondly, we have the following estimate on the norm of $T_r(f)$.

\prop\tfbndprop{$T_r(f)$ is a bounded operator on $\hil$.  Furthermore,
$$
\norm{T_r(f)} \le C \sum_{\alpha,\beta} r^{-(|\alpha|+|\beta|)/2}
\norm{f_{\alpha\beta}}_0.  \num
$$}

\noindent
In particular, a super Toeplitz operator is bounded. We let
${\cal T}_r(\scd)$ denote the $\bC^*$-algebra generated by all super
Toeplitz operators.

The above proposition follows directly from the following lemmas and
proposition.   To simplify the notation, in the rest of the paper we will
suppress the subscript $r$ in $\norm\cdot_r$.

\lemma\fnorm{For $\psi,\phi\in\hil$, and $g\in B^\infty(\cd)$
(an ordinary function) we have
$$
\left| \int_\scd \ol{\psi(Z)} g(z) \phi(Z) \dmu Z \right| \le
\norm{g}_\infty \; \norm{\psi} \; \norm{\phi}.  \num
$$}

\medskip\noindent
{\it Proof.}
Because of \posipthm, we can view $(\cdot,\cdot)_r$ as
an inner product on the space of functions which are holomorphic
only in the odd coordinates.  Thus we have
$$
\left| \int_\scd \ol{\psi(Z)} g(z)\phi(Z)\dmu Z \right|=
|(\psi, g\phi)_r|.
\num
$$
By the Schwarz inequality,
$$
|(\psi, g \phi)_r|\le \norm\psi \; \left\{ \int_\scd
|g(z)|^2\ol{\phi(Z)} \phi(Z) \dmu Z \right\}^{1/2}.  \num
$$
Because $\ol{\phi(Z)} \phi(Z) \dmu Z$ is a positive measure, we can extract
the sup norm of $g(z)$, giving
$$
\left| \int_\scd \ol{\psi(Z)} g(Z) \phi(Z) \dmu Z \right| \le
\norm{g}_\infty
\; \norm\psi \; \norm\phi.  \qquad\square  \num
$$

\lemma\qnorm{For any odd generator $\theta_\mu$,
$$
\norm {T_r(\theta_\mu)} \le Cr^{-1/2},  \num
$$
for $r$ sufficiently large.}

\medskip\noindent
\qnorm\ will be proven in Section \possec.

\prop\fqnormext{For $\psi,\phi\in\hil$, and $f\in B^\infty(\scd)$,
we have
$$
\left| \int_\scd \ol{\psi(Z)} f(Z) \phi(Z) \dmu Z \right| \le  C
\sum_{\alpha,\beta} r^{-(|\alpha|+|\beta|)/2} \;
\norm{f_{\alpha\beta}}_\infty\; \norm\psi \;\norm\phi.  \num
$$
In particular,
$$
\left| \int_\scd \ol{\psi(Z)} f(Z) \phi(Z) \dmu Z \right| \le  C \norm{f}_0
\; \norm\psi \;\norm\phi,  \num
$$
where $\norm\cdot_0$ is the norm defined in \normt.}

\medskip\noindent{\it Proof.}
The statement follows immediately from \fnorm\ and \qnorm.
$\quad\square$

\subsec
To conclude this section, we make the statement that the map $B^\infty
(\scd)\to {\cal T}_r(\scd)$, given by $T_r$, is a deformation quantization.
This statement consists of the following theorems, which will be proven
in Section V.

\thm\trflim{For $f\in B^\infty(\scd)$ bounded, we have
$$
\lim_{r\to\infty} \norm{T_r(f)} = \norm{f_{00}}_0.  \num
$$}

In other words, the classical limit wipes out the fermions. This is
not surprising as fermions do not exist in classical mechanics.

\thm\tftglim{For $f,g\in B^\infty(\scd)$, where the components
$f_{\alpha\beta}$ are compactly supported, there is a constant
$C = C(f,g)$, such that
$$
\biggl\Vert T_r(f)T_r(g) - T_r(fg) + {1\over r}  \sum_{\mu,\nu}
(-1)^{\epsilon_\mu p(f)} T_r\Bigl(P_{\mu\nu}
{\del f\over \del Z_\nu} {\del g\over \del \ol
Z_\mu}\Bigr) \biggr\Vert_r \le Cr^{-2},  \num
$$
for $r$ sufficiently large.}

As a consequence of this theorem, we conclude that ${\cal T}_r(\scd)$ is a
quantum deformation of the Poisson algebra $B^\infty(\scd)$, with
$r\inv$ playing the role of Planck's constant. The assumption that $f$
has compact support is certainly not optimal, but some kind of decay
of at least one symbol at the boundary is clearly needed in our proof.
On the other hand, it is easy to verify that the estimate holds for
any polynomial $f$ and $g$.

\thm\commlim{Under the assumptions of \tftglim,
$$
\Bigl\Vert r \bigl[ T_r(f), T_r(g) \bigr] + T_r\bigl(\{f,g\}\bigr)
\Bigr\Vert_r \le Cr^{-1},  \num
$$
for $r$ sufficiently large.}

\medskip\noindent{\it Proof.}
The proof follows immediately from \tftglim\ and from the definition
\spbdef\ of the super Poisson bracket.  $\quad\square$

\section\possec{Positivity and other properties}

\newsubsec
Theorem VII.1 will be proven after two lemmas are established below.

\lemma\bpluslemma{
\item{(i)} $B_+$ is a multiplicative cone.
\item{(ii)} $\exp B_+ \in B_+$.
\item{(iii)} For $g\in B_+$ nilpotent (i.e. $g$ contains no
term which involves only the even variables), $(1+g)^\lambda \in B_+$ for
every $\lambda > n_1$.}

\medskip\noindent{\it Proof.}
Property (i) follows from the fact that
$$
\bar f f \bar g g = (-1)^{p(f)p(\bar g)} \bar f \bar g f g = \ol{fg} fg.  \num
$$
For (ii) we see that for $f \in B_+$,
$$
\exp f = \sum_{n\ge 0}  {1 \over n!} f^n ,  \num
$$
which is in $B_+$ by (i).  For (iii) we note that
$$
(1+g)^\lambda = \sum_{l=0}^{n_1} {\lambda(\lambda-1) \ldots
(\lambda - l + 1)\over l!} g^l. \qquad\square\num
$$

\lemma\kerbplemma{For the matrix superdomains,
$$
N(Z,Z)^\lambda  \in B_+,  \ref{\kerbpo}
$$
for $\lambda \ge n_1$.}

\medskip\noindent{\it Proof.}
Using the properties of the Berezinian we can rewrite
$$
N(Z,Z) = \det(I_m -zz\star - \theta\theta\star).  \num
$$
If we let $X = (I_m - zz\star)^{-1/2}$, then
$$
N(Z,Z)^\lambda = \det X^{-2\lambda}
\det (I_m - X\theta\theta\star X)^\lambda.  \num
$$
The first factor on the right-hand side is clearly in $B_+$.
Since $B_+$ is a multiplicative cone, and because of item (iii) of
\bpluslemma, we will be done if we can show that
$$
\det (I_m - X\theta\theta\star X) \in 1 + B_+.  \ref{\detitt}
$$

To prove \detitt, we make use of the fact that for any square matrix $A$,
$$
\det(I-A) = 1 + \sum_{n=1}^\infty {(-1)^n\over n!} \tr (\wedge^n A).  \num
$$
Now, for an odd matrix $\eta$,
$$
\eqalign{\tr (\wedge^n \eta\eta\star)
&=  (-1)^{n(n+1)/2} \tr \bigl[(\wedge^n \eta\star)
(\wedge^n\eta)\bigr] \cr
&= (-1)^{n^2} \tr \bigl[(\wedge^n \eta)\star (\wedge^n \eta)\bigr].  \cr}
\num
$$
Applying this to \detitt\ we find
$$
\det (I_m - X\theta\theta\star X) =  1 + \sum_{n=1}^\infty {1 \over n!}\tr
\bigl[(\wedge^n X\theta)\star (\wedge^n X\theta)\bigr].  \num
$$
Since $\tr A\star A$ is clearly in $B_+$ for any matrix of functions
in $B_\ast^\infty(\scd)$, this completes the proof. $\quad\square$

\bigskip\noindent{\it Proof of \posipthm.}
{}From \bercon\ it follows that $\int_\scd g \>dZ \ge 0$, for $g\in B_+$
such that $g_\alpha(z)$ is integrable when $\alpha=(1,1,\ldots, 1)$.
Thus \bpluslemma\ (i) and \kerbplemma\ establish that $(\cdot,\cdot)_r$
is non-negative for $r$ sufficiently large.

It remains to show that the form is strictly positive.  Suppose that
there exists $f\in B^\infty_\ast(\scd)$ such that
$$
\int_\scd \ol{f(Z)} f(Z) \det(I_m - zz\star - \theta\theta\star)^\lambda
dZ = 0. \ref{\ffzero}
$$
If we change variables $\theta\to\theta' = (I_m-zz\star)^{-1/2}\theta$,
then this becomes
$$
\int_\scd \ol{f(Z')} f(Z') \det(I_m - zz\star)^{\lambda -q}
\det(I_m - \theta'{\theta'}\star)^\lambda dZ'.  \num
$$
We now perform the integral over $z$.  Since the measure on $z$
is strictly positive we see that the existence of
an $f$ satisfying \ffzero\ is equivalent to the existence of
$g\in \bigwedge(\bC^{mq})$ such that
$$
\int_\scd \ol{g(\theta)} g(\theta) \det(I_m - \theta\theta\star)^\lambda
d\theta = 0. \ref{\ggzero}
$$

We can assume that $g$ is a homogeneous polynomial of degree $k$,
since homogeneous polynomials of different degree will be orthogonal.
We make the expansion $g(\theta) = \sum_{|\alpha|=k} x_\alpha \theta^\alpha$,
where the sum ranges over multi-indices of length $k$.
If we let $A$ be the matrix
$$
A_{\alpha\beta} = \int \ol{\theta^\alpha} \theta^\beta \det(I_m -
\theta\theta\star)^\lambda d\theta, \num
$$
then \ggzero\ is the statement that $x\star A x =0$.
Consider the leading order of the expansion of $A$ in powers of
$\lambda$:
$$
\eqalign{A_{\alpha\beta} &= \int \ol{\theta^\alpha} \theta^\beta
e^{\lambda \tr \theta\star\theta}d\theta + O(\lambda^{mq-k-1})\cr
&= \int \ol{\theta^\alpha} \theta^\beta
{(\lambda \tr \theta\star\theta)^{mq-k}\over (mq-k)!} d\theta
+ O(\lambda^{mq-k-1}) \cr
&= \lambda^{mq-k} [\delta_{\alpha\beta} + O(1/\lambda)]. \cr} \num
$$
We conclude that for $\lambda$ sufficiently large, $A$ is strictly
positive definite.  Hence $x\star A x=0$ implies $x=0$, i.e. \ggzero\
implies $g=0$. $\quad\square$

\subsec
In this subsection we establish some facts concerning integration over purely
odd matrices.  These facts will be used to prove the remaining technical
assumptions of Section \qntsec\ in the next subsection.
\lemma\colint{Let $\eta$ represent an $m\times 1$ column vector of odd
variables and  let $S^k_m$ denote the set of ordered subsets of
$\{1,\ldots,m\}$ of cardinality $k$.  For $\alpha \in S^k_m$ and $\beta\in
S^l_m$,
$$
\int \ol{\eta_{\alpha_1}\ldots\eta_{\alpha_k}} \> \eta_{\beta_1}
\ldots \eta_{\beta_l}\> (1 - \eta\star \eta)^\lambda d\eta
= \epsilon_{\alpha\beta} {\Gamma(\lambda+1)\over \Gamma(\lambda -
m + k +1)}\;, \num
$$
where $\epsilon_{\alpha\beta}= 0$ unless $\alpha$ is a permutation of
$\beta$ and in this case is given by the sign of the relative permutation.}

\medskip\noindent{\it Proof.}
It is clear that $\beta$ must be a permutation of $\alpha$ for the integral to
be nonzero, since $(1 - \eta\star\eta)^\lambda$ contains only pairs of the
form $\bar\eta_j\eta_j$.  By permuting
the set $\beta$ into the set $\alpha$ and keeping track of the sign, we find
$$
\int \ol{\eta_{\alpha_1}\ldots\eta_{\alpha_k}} \> \eta_{\beta_1}
\ldots \eta_{\beta_l}\> (1 - \eta\star \eta)^\lambda d\eta
= \epsilon_{\alpha\beta}  \int
\ol{\eta_{\alpha_1}\ldots\eta_{\alpha_k}} \> \eta_{\alpha_1}
\ldots \eta_{\alpha_k}\> (1 - \eta\star \eta)^\lambda d\eta . \num
$$
Now we can simply compute
$$
\eqalign{&\int \ol{\eta_{\alpha_1}\ldots\eta_{\alpha_k}} \>
\eta_{\alpha_1}
\ldots \eta_{\alpha_k}\> (1 - \eta\star \eta)^\lambda d\eta\cr
&\qquad= {\Gamma(\lambda+1) \over \Gamma(\lambda - m +k+1)
(m-k)!}
\int \ol{\eta_{\alpha_1}\ldots\eta_{\alpha_k}} \> \eta_{\alpha_1}
\ldots \eta_{\alpha_k} (\eta\star\eta)^{m-k} d\eta \cr
&\qquad= {\Gamma(\lambda+1) \over \Gamma(\lambda - m +k+1)}.
\qquad\square\cr} \num
$$

\lemma\fermint{For $m\times q$ odd matrices $\theta$,
we have
$$
\int \det(I_m -  \theta\theta\star)^\lambda  d\theta =  \prod_{0\le k\le
m-1}  {\Gamma(\lambda-k+q)\over \Gamma(\lambda-k)},  \num
$$
which behaves as $\lambda^{mq}$ for $\lambda\to\infty$.}

\medskip\noindent{\it Proof.}
Decompose $\theta$ into $(\theta',\rho)$, where $\rho$ is the
last column of $\theta$.  We have
$$
I_m - \theta\theta\star = I_m - \theta'{\theta'}\star - \rho\rho\star.
\num
$$
We next define $\omega = (I_m - \theta'{\theta'}\star)^{-1/2}\rho$, so that
$$
I_m - \theta\theta\star = (I_m - \theta'{\theta'}\star)
(I_m - \omega\omega\star).  \num
$$
The change of variables from $\rho$ to $\omega$ gives
$$
\int \det(I_m -  \theta\theta\star)^\lambda  d\theta =
\int \det(I_m -  \theta'{\theta'}\star)^{\lambda-1} d\theta'
\int \det(I_m -  \omega\omega\star)^\lambda  d\omega.  \num
$$

Applying this procedure recursively,
and using the fact that
$$
\det (I_m - \omega\omega\star) = (1 - \omega\star\omega)\inv, \num
$$
we get
$$
\int \det(I_m - \theta\theta\star)^\lambda  d\theta =
\prod_{0\le k\le m-1} \int (1 - \omega\star\omega)^{-(\lambda-k)}
d\omega.  \num
$$
The result then follows from \colint\ with $\alpha=\beta=\emptyset$.
$\quad\square$

\lemma\minorint{Let $a$ be an invertible $m \times m$ ordinary matrix, and
let $\eta$ represent an $m\times 1$ column vector of odd variables.   For
$\alpha \in S^k_m$ and $\beta\in S^l_m$ (the sets defined in \colint), we have
the
integral formula:
$$
\eqalign{&\int \ol{\eta_{\alpha_1} \ldots \eta_{\alpha_k}}
\eta_{\beta_1} \ldots \eta_{\beta_l}
\det(aa\star - \eta\eta\star)^\lambda d\eta \cr
&\qquad=  {\delta_{kl} \Gamma(\lambda+1)\over \Gamma(\lambda-m+k+1)}
\det (aa\star)^{\lambda-1} \det\nolimits_{\beta\alpha}(aa\star), \cr}
\ref{\etaint}
$$
where $\det_{\beta\alpha}$ is the determinant minor taken over the rows
$\beta$ and columns $\alpha$.}

\medskip\noindent{\it Proof.}
The fact that $k$ must be equal to $l$ is clear.  Let $\eta = a\omega$.  Then
$$
\det(aa\star - \eta\eta\star) = (1 - \omega\star\omega) \det(aa\star),
\num
$$
and the measure transforms to $d\omega = \det(aa\star) d\eta$.  Thus
under the change of variables the left-hand side of \etaint\ becomes
$$
 \det(aa\star)^{\lambda-1} \int \ol{(a\omega)_{\alpha_1}\ldots
(a\omega)_{\alpha_k}} (a\omega)_{\beta_1} \ldots (a\omega)_{\beta_l}
(1 - \omega\star\omega)^\lambda d\omega.  \num
$$
We now apply \colint\ to perform the integration over $\omega$.  The
result is
$$
\det(aa\star)^{\lambda-1} {\Gamma(\lambda+1)\over
\Gamma(\lambda-m+k+1)}  \sum_{\mu,\nu \in S^k_m}
\epsilon_{\mu\nu}
\bar a_{\alpha_1\mu_1} \ldots \bar a_{\alpha_k\mu_k}
a_{\beta_1\nu_1} \ldots a_{\beta_k\nu_k}. \ref{\munusum}
$$
The sum in \munusum\ can be rewritten as
$$
\eqalign{&\sum_{\mu\in S^k_m} \sum_{\sigma\in S_k} \epsilon(\sigma)
\bar a_{\alpha_1\mu_1} \ldots \bar a_{\alpha_k\mu_k}
a_{\beta_1\mu_{\sigma(1)}} \ldots a_{\beta_k\nu_{\sigma(k)}} \cr
&\qquad=
\sum_{\mu\in S^k_m} \sum_{\sigma\in S_k} \epsilon(\sigma)
\bar a_{\alpha_{\sigma(1)}\mu_1} \ldots \bar
a_{\alpha_{\sigma(k)}\mu_k}  a_{\beta_1\mu_1} \ldots a_{\beta_k\nu_k}
\cr
&\qquad= \sum_{\sigma\in S_k} \epsilon(\sigma)
(aa\star)_{\beta_1\alpha_{\sigma(1)}} \ldots
(aa\star)_{\beta_k\alpha_{\sigma(k)}}  , \cr} \num
$$
where $S_k$ denotes the set of permutations of $\{1,\ldots,k\}$ and
$\epsilon(\sigma)$ is the sign of the permutation $\sigma$.
This final sum over $\sigma$ is precisely the definition of the determinant
minor $\det_{\beta\alpha}(aa\star)$.  $\quad\square$

\subsec
We turn now to the proofs of Proposition VII.3 and Lemma VII.11.
We can in fact replace Proposition VII.3 by the following,
stronger statement.
\prop\lrprop{
$$
\int N(Z,Z)^{r-p} d^{2n_1}\theta
= C_r r^{n_1} \Delta(z,z)^{r-p_0},  \num
$$
where $\Delta(z,z) := \det(I_m - zz\star)$ is the triple
determinant of the underlying domain and where
$$
C_r = \prod_{0\le k\le
m-1}  {\Gamma(r-p_0-k)\over \Gamma(r-p-k)}. \num
$$}

\medskip\noindent{\it Proof.}
The function $N(Z,Z)$ has the form
$$
\eqalign{N(Z,Z) &= \det(I_m - zz\star -
\theta\theta\star) \cr
&= \det(I_m -  zz\star) \det\bigl(I_m - (I_m - zz\star)\inv
\theta\theta\star\bigr).  \cr}   \num
$$
By changing variables $\theta \to \theta'= (I_m - zz\star)^{-1/2}\theta$,
we obtain
$$
\eqalign{\int N(Z,Z)^{r-p} d\theta &= \det(I_m - zz\star)^{r-p_0+q}
\int \det\bigl(I_m - (I_m - zz\star)\inv
\theta\theta\star\bigr)^{r-p}  d^{2mq}\theta \cr
&= \det(I_m - zz\star)^{r-p_0} \int
\det(I_m -  \theta'{\theta'}\star)^{r-p}  d^{2mq}\theta'.\cr} \ref{\idmurz}
$$
The proof follows from \fermint.  $\quad\square$

\bigskip\noindent{\it Proof of Lemma VII.11.}
We need to compare $\norm\phi$ to $\norm{\theta_{ij}\phi}$.   To do this
we start by integrating over all of the odd variables except for the $j$-th
column.  Denote the $j$-th column by $\eta$, so that $\theta_{ij} = \eta_i$,
and denote the remaining odd variables by $\theta'$.  Let $\lambda$ denote
$r-p$.  The integral over $\theta'$ is
$$
\eqalign{\norm\phi^2 &= \Lambda_r \int_\scd \ol\phi \phi
\det(I - zz\star - \theta'{\theta'}\star - \eta\eta\star)^\lambda
\>d\theta'\> d\eta\>dz\cr
&= \Lambda_r \int_\scd \int \Psi(z,\eta) \det(I - zz\star -
\eta\eta\star)^\lambda d\eta\> dz,\cr}  \ref{\tpint}
$$
where
$$
\Psi(z,\eta) := \int \ol\phi \phi \det\bigl(I - (I-zz\star - \eta\eta\star)\inv
\theta'{\theta'}\star \bigr)^\lambda d\theta'.  \num
$$
For $\norm{\theta_{ij}\phi}$ we obtain
\tpint\ with $\Psi$ replaced by $\Psi \bar\eta_i \eta_i$.

Let $\hat S^k_m$ denote the set $\{\alpha\in S^k_m:
\alpha_1<\ldots<\alpha_k\}$.
We can decompose the function $\Psi$ uniquely into
$$
\Psi(z,\eta) = \sum_{k=0}^m \sum_{\mu,\nu \in \hat S^k_m}
\Psi_{\mu\nu}(z)
\ol{\eta_{\mu_1}\ldots \eta_{\mu_k}} \eta_{\nu_1} \ldots \eta_{\nu_k}.
\num
$$
By \minorint\ the norm of $\phi$ is given by
$$
\norm\phi^2  = \Lambda_r \sum_{k=0}^m {\Gamma(\lambda+1)\over
\Gamma(\lambda - m+k+1)} \sum_{\mu,\nu \in S^k_m} \int_\cd
\Psi_{\mu\nu}(z) \det\nolimits_{\mu\nu}(I-zz\star)  \det(I-
zz\star)^{\lambda-1}  dz, \ref{\nphi}
$$
and the norm of $\norm{\theta_{ij}\phi}$ is
$$
\eqalign{\norm{\theta_{ij}\phi}^2 &= \Lambda_r \sum_{k=0}^{m-1}
{\Gamma(\lambda+1)\over \Gamma(\lambda - m+k+2)} \cr
&\qquad\times \sum_{\mu,\nu \in S^k_m} \int_\cd
\Psi_{\mu\nu}(z) \det\nolimits_{\mu+\{i\},\nu+\{i\}}(I-zz\star)  \det(I-
zz\star)^{\lambda-1}  dz, \cr}\ref{\nthphi}
$$
where $\mu+\{i\}$ denotes the sequence $\mu$ with $i$ inserted in the
appropriate location.

For the rest of the proof we will consider a fixed value of $k$ and work
pointwise in $z$.  Note that the difference between the multiplier outside the
integral in \nphi\ and that of \nthphi\ is $(\lambda - m +k+1)\inv$.
Because the factor $\det(I_m-ZZ\star)^\lambda$ of the measure is in $B_+$,
it follows that the function $\Psi(z,\eta)$ is also in $B_+$.
Thus $\Psi$ must have the form
$$
\Psi = \sum_j \ol{X_j(z,\eta)} X_j(z,\eta), \num
$$
where $X_j\in \bigwedge^k (\bC^m)$ can be written
$$
X_j(z,\eta) = \sum_{\nu \in \hat S^k_m} X_{j,\nu}(z) \eta_{\nu_1} \ldots
\eta_{\nu_k}.  \num
$$

Let $\bC^m$ be given the inner product $(u,v) = u\star(I_m- zz\star)v$.
The natural extension of an inner product to an exterior algebra is simply
the determinant minor, i.e.
$$
\bigl(\eta_{\mu_1}\ldots\eta_{\mu_k},
\eta_{\mu_1}\ldots\eta_{\mu_k}\bigr) = \det\nolimits_{\mu\nu}(I_m-zz\star).
\num
$$
This means that we can write the integrand of \nphi\ as
$$
\det(I_m - zz\star)^{\lambda-1} \sum_j
\norm{X_j}^2_{\wedge^k(\bC^m)}, \num
$$
and the integrand of \nthphi\ as
$$
\det(I_m - zz\star)^{\lambda-1} \sum_j
\norm{\eta_i X_j}^2_{\wedge^{k+1}(\bC^m)}. \num
$$

The problem then reduces to computing the norm of the operator $\sigma_i:
\bigwedge^k(\bC^m) \to \bigwedge^{k+1}(\bC^m)$ which maps $X\mapsto
\eta_iX$.   Let $\omega = a \eta$ where $aa\star = (I_m - zz\star)$.  Then
the $\omega$'s generate an orthonormal basis for $\bigwedge(\bC^m)$.
The Hilbert-Schmidt norm of $\sigma_i$ is easily computed:
$$
\eqalign{\norm{\sigma_i}^2_2 &= \sum_{\mu\in \hat S^k_m}
\norm{\eta_i \omega_{\mu_1}\ldots
\omega_{\mu_k}}^2_{\wedge^{k+1}(\bC^m)} \cr
&= \sum_{\mu\in \hat S^k_m} \sum_{l\notin\mu} \bar a_{il} a_{il}\cr
&= {m-1 \choose k} (I_m-zz\star)_{ii}\cr
&\le {m-1 \choose k}.\cr}  \num
$$
Since this estimate is independent of $z$, we can use it inside the integral in
\nthphi.  We conclude that the $k$-th summand of \nthphi\ can be bounded
by
$$
{m-1\choose k} {1\over \lambda - m+k+1}, \num
$$
times the $k$-th summand of \nphi. $\quad\square$.

\section\qdefsec{Proof of deformation estimates}

\newsubsec
In this section we prove \trflim\ and \tftglim\ for a generic Cartan
superdomain of type I--III.

\bigskip\noindent
{\it Proof of \trflim.}
{}From \fnorm\ and \qnorm\ we have
$$
\norm{T_r(f)} \le \norm{f_{00}}_\infty + O(r^{-1/2}),  \num
$$
as $r\to\infty$, i.e. $\lim \sup_{r\to\infty} \norm{T_r(f)} \le
\norm{f_{00}}_\infty$.  We will show below that
$$
\norm{f_{00}}_\infty \le \norm{T_r(f)} + o(1),  \ref{\liminf}
$$
as $r\to\infty$, i.e. $\lim \inf_{r\to\infty} \norm{T_r(f)} \ge
\norm{f_{00}}_\infty$, and the claim will follow.

To prove \liminf, we set $Z = (z,0)$ and write
$$
\eqalign{f(Z) &= f_{00}(z) \cr
&= (\phi_0, T_r(f\circ\gz) \phi_0) + \Bigl\{f_{00}(z) - \int_\scd f(\gz(W))
\dmu W \Bigr\}, \cr} \num
$$
where $\phi_0 = 1$ is the vacuum element.  Using \trfcg,
we rewrite the above equation as
$$
\eqalign{f_{00}(z) &=  (\phi_0, U(\gz)\inv T_r(f) U(\gz) \phi_0) \cr
&\quad + \Bigl\{f_{00}(z) - \int_\scd f_{00}(w')
\dmu W \Bigr\}  \cr
&\quad +  \int_\scd  \bigl[f(\gz(W)) - f_{00}(w')\bigr] \dmu W, \cr}
\ref{\fooz}
$$
where $(w',\eta') := \gz(W)$.
The first term in \fooz\ can be bounded by
$\norm{T_r(f)}$, as $U(\gz)$ is unitary.  Using \fintprop, we can
apply the proof of Theorem 2.1 in
[6] to show that the second term is $o(1)$ uniformly in $z$,
as $r\to\infty$.  For the
third term, we use \fqnormext\ to bound
$$
\left| \int_\scd  \bigl[f(\gz(W)) - f_{00}(w')\bigr] \dmu W  \right| \le C
\sum_{\alpha,\beta, \> |\alpha|=|\beta|\ne 0} r^{-(|\alpha|+|\beta|)/2}
\norm{(f\circ\gz)_{\alpha\beta}}_\infty,
\num
$$
and the claim follows.  $\quad\square$

\subsec
In this subsection, we give two lemmas which will be needed
for the proof of \tftglim.
For the following lemma and its proof we extend the norm $\norm\cdot_0$
to supermatrices by taking the supremum of the norms of the elements of the
matrix.  We denote by $\gamma^{(k)}(W)$ the $k$-th complex derivative
of $\gamma(W)$.
\lemma\gzbnd{
For each $k$, there exist constants $s,s' > 0$ such that
$$
\left\Vert \gz^{(k)}(W)\right\Vert_0  \le C
\Delta(w,w)^{-s}  \Delta(z,z)^{-s'} , \num
$$
where $\Delta(z,z)$ denotes the triple determinant of
the underlying Cartan domain.}

\medskip\noindent{\it Proof.}
For type I superdomains, the first complex derivative $\gz'(W)$
was computed in the proof of \gpzpropi\ to be
$$
\gz'(W) = \Bigl[ (WB\star + A\star)^T \otimes (CW +D)
\Bigr]\inv,  \ref{\gpw}
$$
where $A,B,C,D$ are the matrix blocks of $\gz$.
For types II and III the computation is essentially the same, although
the tensor product will be replaced by some partially symmetrized
or antisymmetrized tensor product.  This will not affect the bounds,
so we proceed to analyze \gpw.

For the following discussion, we abuse notation slightly by letting
$\Vert A \Vert_0$, for a supermatrix $A$, denote the supremum of
the $\Vert \cdot\Vert_0$ of all the entries.  Each
further derivative of \gpw\ will involve an extra factor of
$(ZB\star + A\star)\inv$ or $(CZ+D)\inv$, times entries of
$B\star$ and $C$, respectively.  For types II and III there
will be extra factors of two, but this will not make a difference.
By a conservative estimate we have
$$
\left\Vert \gz^{(k)}(W) \right\Vert_0
\le K \Bigl[\norm B_0 \norm C_0\Bigr]^{k-1}
\Bigl[\norm{(WB\star + A\star)\inv}_0
\norm{(CW+D)\inv}_0\Bigr]^k, \num
$$
where $K$ is some constant.

For the matrices $B$ and $C$ we have, by virtue of the conditions
\grelns\ the bounds $\norm B_0 \le \norm A_0$ and
$\norm C_0 \le \norm D_0$.
Now, for all domains, $A$ and $D$ satisfy the relations \adsat,
which implies $\norm A_0 \le \norm{(I_m - ZZ\star)\inv}_0$ and
$\norm D_0 \le \norm{(\inq - Z\star Z)\inv}_0$.  Furthermore,
up to a constant matrix, $CW + D = D\inv (\inq + Z\star W)$
and $WB\star + A\star = (I_m + WZ\star) A\star$.  Thus the proof
will be finished if we can establish a bound
$$
\norm{(\inq + Z\star W)\inv}_0 \le K \Delta(w,w)^{-s}
\Delta(z,z)^{-s'}  \num
$$
(the case of $(I_m + WZ\star)\inv$ is similar enough that it need not
be dealt with separately).

To make this bound, we observe that
$$
\eqalign{(\inq + Z\star W)\inv &= \pmatrix{I_n& - (I_n+z\star w)\inv
z\star \eta\cr
-(I_q + \eta\theta\star)\inv\theta\star\eta& I_m\cr} \cr
&\qquad\times
\pmatrix{I_n + z\star(I_m + \eta\theta\star)\inv
w &0\cr 0& I_q + \theta\star(I_m + wz\star)\inv\eta\cr}\inv. \cr} \num
$$
It is clear that the only divergent matrix elements in this
expression come from the matrix elements of $(I_n + z\star w)\inv$.
This is precisely the divergent factor in the case of ordinary
domains, and so the result follows from the proof of
[6], Lemma 3.2 (ii).  $\quad\square$

\lemma\uvphi{For $u,v \in B^\infty(\scd)$, and $\phi\in\hil$,
we have
$$
\eqalign{&\left| \int_\scd u(W) v(W) \phi(W) \dmu W \right| \cr
&\qquad\le C\norm\phi \>\norm v_0 \sum_{\alpha,\beta}
r^{-(|\alpha|+|\beta|)/2}
\left\{\int_\scd |u_{\alpha\beta}(w)|^2  \dmu W
\right\}^{1/2}.\cr} \num
$$}
{\it Proof.}
We write
$$
\eqalign{\left| \int_\scd u(W) v(W) \phi(W) \dmu W \right|  &\le
\sum_{\alpha,\beta,\rho,\delta} \left| \int_\scd u_{\alpha\beta}(w)
\bar\eta^\alpha \eta^\beta v_{\rho\delta}(w) \bar\eta^\rho
\eta^\delta  \phi(W) \dmu W \right| \cr
&= \sum_{\alpha,\beta,\rho,\delta}
\Bigl|\bigl(\bar u_{\alpha\beta}(w) \eta^\alpha \eta^\rho ,
v_{\rho\delta}(w)  \eta^\beta \eta^\delta  \phi(W)\bigr)\Bigr|.\cr}  \num
$$
By virtue of \posipthm\ we can apply the Schwarz inequality
to this expression to obtain
$$
\left| \int_\scd u(W) v(W) \phi(W) \dmu W \right|  \le
\sum_{\alpha,\beta,\rho,\delta}
\norm{v_{\rho\delta}}_\infty  \;\norm{\bar u_{\alpha\beta}(w) \eta^\alpha
\eta^\rho}  \; \norm{\eta^\beta \eta^\delta  \phi(W)}.  \num
$$
By \qnorm\ we then have
$$
\eqalign{\left| \int_\scd u(W) v(W) \phi(W) \dmu W \right|
&\le C\sum_{\alpha,\beta,\rho,\delta} r^{-(|\alpha| + |\beta| + |\rho|
+|\delta|)/2}
\norm{v_{\rho\delta}}_\infty  \;\norm{\bar u_{\alpha\beta}}
 \; \norm\phi \cr
&\le C\norm\phi\;\norm{v}_0 \sum_{\alpha,\beta} r^{(|\alpha| + |\beta|)/2}
\norm{\bar u_{\alpha\beta}}. \qquad\square\cr} \num
$$

\subsec
{\it Proof of \tftglim.}
Our procedure will be to expand
$$
\bigl(\phi,T_r(f) T_r(g) \psi\bigr)  = \int_{\scd\times\scd} \ol{\phi(Z)}
f(Z)
\rker
ZX g(X) \psi(X) \dmu Z \dmu X,   \ref{\tftgint}
$$
where $\psi,\phi\in\hil$, $f,g\in B^\infty(\scd)$, in a power series
in $r$ [12].  We make the substitution $X = \gz(W)$, and use the
transformation properties of the Bergman kernel to rewrite \tftgint\ as
$$
\eqalign{&\bigl(\phi,T_r(f) T_r(g) \psi\bigr)\cr
&\quad= \int_{\scd\times\scd} \ol{\phi(Z)} f(Z)
{\rker ZZ \over \rker{\gz(W)}Z} g(\gz(W)) \psi(\gz(W)) \dmu Z \dmu W. \cr}
\num
$$
The next step will be to expand $g(\gz(W))$ in a Taylor series.  We will
need to expand out to order $m$, where $m$ is an integer
such that $m > n_0 + 4$.
The Taylor expansion for superfunctions is:
$$
\eqalignno{g(\gz(W))
= g&(Z) + \sum_{\mu,\kappa}\;\big(W_\kappa \dgz \kappa\mu
\;\del_\mu g(Z)  + \ol{W_\kappa \dgz \kappa\mu}
\;\dbr_\mu g(Z)\big) \cr
&+\half\sum_{\mu,\nu,\kappa,\rho} W_\kappa \dgz \kappa\mu
\;W_\nu \dgz\nu\rho \;\del_\rho \del_\mu g(Z) \cr
&+\half\sum_{\mu,\kappa,\rho} W_\kappa W_\rho
\Gamma_{\rho\kappa\mu}(Z) \del_\mu g(Z) \cr
&+\sum_{\mu,\nu,\kappa,\rho} \ol{W_\kappa \dgz \kappa\mu}  \;
W_\rho \dgz \rho\nu \del_\nu \dbr_\mu g(Z) &\eqalignnum \cr
&+ \half\sum_{\mu,\nu,\kappa,\rho} \ol{W_\kappa
\dgz\kappa\mu}  \;\ol{W_\rho \dgz\rho\nu}
\;\dbr_\nu \dbr_\mu g(Z) \cr
&+\half\sum_{\mu,\kappa,\rho} \ol{W_\kappa W_\rho
\Gamma_{\rho\kappa\mu}(Z)} \;\dbr_\mu g(Z) \cr
&+\hbox{\it  terms of order $3$ through $m-1$}\cr
&+ G(Z,W), \cr}
$$
where $\del_\mu := \del/\del Z_\mu$ and
$$
\Gamma_{\rho\kappa\mu}(Z) := {\del\over\del W_\rho}
{\del\over\del W_\kappa} \gz(W)_\mu  \Bigr|_{W=0},  \num
$$
and the $m$-th order remainder term is given by
$$
G(Z,W) := {1\over (m-1)!} \int_0^1 ds (1-s)^{(m-1)} {d^m\over ds^m}
g(\gz(sW)).   \ref{\remterm}
$$

Denote by $I_{a,b}$ the contribution to the integral from the term in the
expansion of $g$ with $a$ powers of $W$ and $b$ powers of $\ol W$, and let
$R$ denote the contribution of the remainder term.  In evaluating these
terms we will make use of the following facts.  Given a
holomorphic function $\chi$ on $\scd$,
we have remarked before that
$$
\int_\scd   \chi(W) \dmu W =  \chi(0). \ref{\holfact}
$$
Furthermore, using the circular symmetry and \kzwprop\ we obtain
$$
\eqalign{
\int_\scd \ol W_\mu \chi(W) \dmu W &= {\del \chi\over \del W_\mu} (0)
\int_\scd \ol W_\mu W_\mu \dmu W.  \cr
&= {\beta_\mu\over r}
{\del \chi\over \del W_\mu} (0), \cr} \ref{\olwint}
$$
for any $\mu$.

For the lowest order term in the expansion, we have
$$
I_{0,0} = \int_{\scd\times\scd} \ol{\phi(Z)} f(Z) {\rker ZZ\over
\rker{\gz(W)}Z}
g(Z) \psi(\gz(W)) \dmu Z \dmu W.  \num
$$
The integrand is holomorphic in $W$, so we apply \holfact\ to get
$$
\eqalign{I_{0,0} &= \int_{\scd} \ol{\phi(Z)} f(Z) g(Z) \psi(Z) \dmu Z
\cr
&= (\phi,T_r(fg) \psi).  \cr}\num
$$
The same fact \holfact\ also clearly implies that $I_{a,b} = 0$ for $a>b$.

The next nonzero term in the expansion is thus $I_{0,1}$, which is given by
$$
\eqalign{I_{0,1} =  \sum_{\mu,\kappa} \int_{\scd\times\scd} \ol{\phi(Z)}
f(Z) &
\ol{W_\kappa \dgz \kappa\mu}\; \dbr_\mu g(Z)  {\psi(\gz(W))\over
\rker{\gz(W)}Z} \cr
&\rker ZZ \dmu Z \dmu W.\cr} \num
$$
We now apply \olwint, to obtain
$$
\eqalign{I_{0,1} = {1\over r}\sum_{\mu,\kappa}
(-1)^{\epsilon_\kappa(p(g)+\epsilon_\mu)} \beta_\kappa
&\int_\scd \ol{\phi(Z)} f(Z)  \ol{\dgz \kappa\mu}
\;\dbr_\mu g(Z) \cr
&\times \left[ {\del\over \del W_\kappa} \>{\psi(\gz(W))\over
\rker{\gz(W)}Z}\right]_{W=0} \rker ZZ \dmu Z ,\cr}  \num
$$
where the sign arises from the permutation of elements of
the integrand (keeping in mind the fact that $\ol{W\Gamma}
=\ol\Gamma\; \ol W$).  Applying the chain rule gives
$$
\eqalign{I_{0,1} &= {1\over r} \sum_{\mu,\nu,\kappa} (-
1)^{\epsilon_\kappa(p(g)+\epsilon_\mu)} \beta_\kappa\cr
&\qquad\times \int_\scd \ol{\phi(Z)} f(Z)
\ol{\dgz \kappa\mu} \;\dbr_\mu g(Z) \; \dgz \kappa\nu \; \del_\nu \!
\left[ {\psi(Z)\over \rker ZZ}\right] \rker ZZ  \dmu Z  \cr
&= {1\over r} \sum_{\mu,\nu} (-
1)^{(\epsilon_\mu+\epsilon_\nu)p(f)+\epsilon_\nu(p(g)+\epsilon_\mu)}
\cr
&\qquad\times
\int_\scd \ol{\phi(Z)}\; P_{\mu\nu}(Z) f(Z) \>
\dbr_\mu g(Z)
\;\del_\nu\! \left[ {\psi(Z)\over \rker ZZ}\right] \rker ZZ \dmu Z.
\cr}  \num
$$
Noting that
$$
\rker ZZ \dmu Z = N(Z,Z)^{-p} dZ,  \num
$$
we integrate by parts as follows:
$$
\eqalign{I_{0,1} = -& {1\over r} \sum_{\mu,\nu} (-
1)^{\epsilon_\mu p(f) +\epsilon_\nu(\epsilon_\mu+1)}
\int_\scd
\ol{\phi(Z)}\; \del_\nu \!\left[ {P_{\mu\nu}(Z) \over
N(Z,Z)^p}  f(Z)  \dbr_\mu g(Z)  \right]\cr
&\hskip2in \times\psi(Z)  N(Z,Z)^p \dmu Z    \cr
= -& {1\over r} \sum_{\mu,\nu} (-1)^{\epsilon_\mu p(f) +
\epsilon_\nu(\epsilon_\mu+1)} \int_\scd \ol{\phi(Z)}
\; \del_\nu \! \Bigl[ {P_{\mu\nu}(Z) \over
N(Z,Z)^p} \Bigr]  f(Z) \dbr_\mu g(Z) \cr
&\hskip2in \times \psi(Z)  N(Z,Z)^p  \dmu{Z}  \cr
-& {1\over r} \sum_{\mu,\nu} (-1)^{\epsilon_\mu
p(f)}
\int_\scd \ol{\phi(Z)} \;P_{\mu\nu}(Z) \del_\nu
f(Z)  \dbr_\mu g(Z)  \psi(Z)   \dmu Z   \cr
-& {1\over r} \sum_{\mu,\nu} (-
1)^{(\epsilon_\mu+\epsilon_\nu)p(f)} \int_\scd \ol{\phi(Z)}
\;P_{\mu\nu}(Z) f(Z) \del_\nu\dbr_\mu
g(Z)  \psi(Z)  \dmu Z \; .   \cr } \ref{\ioone}
$$
Observe that, as a consequence of the assumption that $r$ is sufficiently
large, no boundary terms are present.  As a consequence of \dsthm,
$$
\sum_{\nu} (-1)^{\epsilon_\nu(\epsilon_\mu+1)}
 \;\del_\nu\! \left[ {P_{\mu\nu}(Z) \over N(Z,Z)^p}\right] = 0.  \num
$$
This leaves two terms in \ioone.

Now consider the term $I_{1,1}$, which is given by
$$
\eqalign{
I_{1,1} = \sum_{\mu,\nu,\kappa,\rho} \int_{\scd\times\scd} \ol{\phi(Z)}
f(Z) &{\rker ZZ\over \rker{\gz(W)}Z}
\ol{W_\kappa \dgz \kappa\mu}  W_\rho \dgz \rho\nu \;\del_\nu
\dbr_\mu g(Z)\cr
&\times \psi(\gz(W)) \dmu Z
\dmu W.  \cr}\num
$$
Using \holfact\ and \olwint, we can perform the
$W$ integration to get
$$
\eqalign{I_{1,1} &= {1\over r} \sum_{\mu,\nu} \int_\scd
\ol{\phi(Z)} f(Z)
P_{\mu\nu}(Z) \del_\nu \dbr_\mu g(Z) \psi(Z)
\dmu Z  \cr
&= {1\over r}\sum_{\mu,\nu} (-
1)^{(\epsilon_\mu+\epsilon_\nu)p(f)} \int_\scd \ol{\phi(Z)}
\;P_{\mu\nu}(Z)  f(Z)
\del_\nu\dbr_\mu g(Z)   \psi(Z)  \dmu Z .   \cr} \num
$$
This exactly cancels the third term in \ioone, so that we
finally obtain
$$
I_{0,1} + I_{1,1} = {1\over r}\sum_{\mu,\nu,\kappa}
(-1)^{\epsilon_\mu p(f)+1} \Bigl(\phi, T_r\bigl(P_{\mu\nu} \>
 \del_\nu f \> \dbr_\mu g\bigr) \psi\Bigr).  \num
$$

All that remains to complete the proof is to bound the other terms as
$r\to\infty$.  Of the remaining second order terms, $I_{2,0}=0$, and
$I_{0,2}$ is given by
$$
\eqalign{
I_{0,2} = \half &\int_{\scd\times\scd} \ol{\phi(Z)} f(Z)
\rker ZZ \rker{\gz(W)}Z\inv \cr
&\qquad\times\biggl[
\sum_{\mu,\nu,\kappa,\rho} \ol{W_\kappa \dgz \kappa\mu} \;
\ol{W_\rho \dgz \rho\nu}
\;\dbr_\nu \dbr_\mu g(Z) \cr
&\qquad+ \sum_{\mu,\kappa,\rho} \ol{W_\kappa W_\rho
\Gamma_{\rho\kappa\mu}(Z)} \;\dbr_\mu g(Z) \biggr]
\psi(\gz(W)) \dmu Z \dmu W.\cr}  \ref{\iotwo}
$$
We want to bound this term for large $r$.  To do this, we first evaluate the
integration over $W$ using the principles of \holfact\ and \olwint.
For this integral we obtain
$$
\eqalign{&\int_\scd \rker{\gz(W)}Z\inv
\ol{W_\kappa W_\rho} \psi(\gz(W)) \dmu W \cr
&\qquad= {1\over 2} \sum_{\mu,\nu} {\del^2 \over \del W_\nu \del W_\mu}
{\psi(\gz(W))\over \rker{\gz(W)}Z}\Big|_{W=0}
\int_\scd \ol{W_\kappa W_\rho} W_\mu
W_\nu \dmu W. \cr} \ref{\wkwrint}
$$

The convergence factor comes from the integral on the right-hand
side of \wkwrint.
We can apply the positivity property of the measure and
the Schwarz inequality to give
$$
\Big|\int_\scd \ol{W_\kappa W_\rho} W_\mu W_\nu \dmu W\Big| \le
\int_\scd \Bigl(\sum_\mu \ol W_\mu W_\mu \Bigr)^2 \dmu W.  \num
$$
Because of \fintprop, we can apply the fact ([6], Lemma 3.1 (ii)) that
$$
{\int_D \bigl(\sum_\mu |w_\mu|^2 \bigr)^{k} \Delta(z,z)^{r-p_0} dz \over
\int_D \Delta(z,z)^{r-p_0} dz} \le Cr^{-k},  \num
$$
together with \fqnormext, to see that
$$
\int_\scd \Bigl(\sum_\mu \ol W_\mu W_\mu \Bigr)^k \dmu W  \le Cr^{-
k}.  \ref{\cfact}
$$

Substituting \wkwrint\ into \iotwo, we
convert the derivatives with respect to $W$ at zero into derivatives
with respect to $Z$ using the chain rule.  We then integrate by parts
to move these derivatives off of the $\psi(Z)$, as in the analysis of
$I_{0,1}$.  These derivatives then act on the expression
$$
f(Z) N(Z,Z)^{-p} \Bigl[ \ol{\gz'(0)_{\kappa\mu}} \>
\ol{\gz'(0)_{\rho\nu}} \>\dbr_\nu \dbr_\mu g(z)
+ \ol{\Gamma_{\rho\kappa\mu}(Z)}\> \dbr_\mu g(Z)\Bigr].  \ref{\deract}
$$
The derivatives of $N, \gz'$, and $\Gamma$ have potential singularities.
In view of \gzbnd\ we can bound the absolute values of the components
of these terms by $\Delta(z,z)^{-s}$ for some integer $s$.
Then, since the supports of the components of the function $f$
are restricted to some compact set $S_f$, we can bound the $\norm\cdot_0$
norm of the derivatives of \deract\ by
$$
C \norm{f}_t \; \norm{g}_t \; \sup_{S_f} \Delta(z,z)^{-s},  \num
$$
for some $t$.
Using this bound in conjunction with \fqnormext, we thus have
$$
|I_{0,2}| \le C_{S_f} r^{-2} \norm{f}_t \;\norm{g}_t\;
\norm\psi\;\norm\phi,  \num
$$
where the $r^{-2}$ comes from the convergence factor \cfact\ and the
constant $C_{S_f}$ depends on $S_f$.

The same reasoning applies to the cases $I_{a,b}$ where $3\le a+ b < m$.
The convergence factor comes from \cfact.  The result is that
$$
|I_{a,b}| \le Cr^{-2} \norm{f}_t \; \norm{g}_t\; \norm\psi\; \norm\phi,
\num
$$
for some $t$ and for $3\le a+b \le m$.

Finally, we turn to the remainder term, which is
$$
R =  \int_{\scd\times\scd} \ol{\phi(Z)} f(Z)
G(Z,W) {\psi(\gz(W)) \over\rker{\gz(W)}Z} \rker ZZ \dmu Z \dmu W.  \num
$$
Note that
$$
\eqalign{{\psi(\gz(W)) \over\rker{\gz(W)}Z} &= \ber \gamma'\ssub Z(W)^r
\ol{\ber \gamma'\ssub Z(0)^r} \psi(\gz(W))  \cr
&= \rker ZZ^{-1/2} \;U(\gamma\inv\ssub Z)\psi(W), \cr}  \num
$$
where $U$ is the projective unitary representation of $\aut$ on $\hil$, and
where we have used the fact that $\gamma'\ssub Z(0)$ is real.  Denote
$U(\gamma\inv\ssub Z)\psi(W)$ by $\psi\ssub Z(W)$, noting that
$\norm{\psi\ssub Z} = \norm \psi$. The remainder term can thus be written
$$
R = \int_{\scd\times\scd} \ol{\phi(Z)} \rker ZZ^{1/2} f(Z)
G(Z,W) \psi\ssub Z(W)   \dmu Z \dmu W.  \ref{\rnowwr}
$$

We can write the components of the function $G(Z,W)$ as
$$
G(Z,W) :=  \sum_{\alpha,\beta,\gamma,\delta}
G_{\alpha\beta\gamma\delta}(z,w) \bar\theta^\alpha \theta^\beta
\bar\eta^\gamma
\eta^\delta,  \num
$$
where the sum is over multi-indices.
For some positive integers $s, s'$, we claim that we have the bound
$$
\sup_z |G_{\alpha\beta\gamma\delta}(z,w)| \le C\norm g_t\;
|w|^{m-|\gamma|-|\delta|} \Delta(w,w)^{-s} \Delta(z,z)^{-s'}  ,
\ref{\biggbnd}
$$
where $|w|^2 := \sum_\mu |w_\mu|^2$.  This
may be established as follows.
Consider the definition of $G(Z,W)$, equation \remterm,
which involves taking $m$ derivatives.
Each derivative with respect to $s$ in \remterm\ brings
out a factor of $W$, since
only the combination $sW$ appears in the definition.
This accounts for the $|w|^{m-|\gamma|-|\delta|}$ appearing in
\biggbnd.
The statement  then follows from \gzbnd.

\bigskip
Applying \uvphi\ to the $Z$ integrations in \rnowwr,
we obtain
$$
|R| \le C \norm{f}_0 \;\norm\phi \sum_{\alpha,\beta} r^{-(|\alpha|
+|\beta|)/2} \Bigl\{ \int_\scd |u_{\alpha\beta}(z)|^2 X_{S_f}(z) \dmu Z
\Bigr\}^{1/2},  \ref{\zintbnd}
$$
where $u(Z)$ is the function
$$
u(Z) = \rker ZZ^{1/2} \int_\scd  G(Z,W) \psi\ssub Z(W) \dmu W, \num
$$
and $X_{S_f}$ is the characteristic function of the compact set
$S_f$ in which the components of $f$ are supported.
Now, to bound the components of $u$, we apply \uvphi\ to the
$W$ integration using the bound \biggbnd.  In this way we find
$$
\eqalign{|u_{\alpha\beta}(z)| &\le C \norm\psi \;
\norm{g}_t \;\Delta(z,z)^{s'}
\norm{\rker ZZ^{1/2}}_0 \cr
&\qquad\times \sum_{\gamma\delta} r^{-(|\gamma|+|\delta|)/2}
\Bigl[ \int_\scd |w|^{2(m-|\gamma|-|\delta|)}  \Delta(w,w)^{-2s}
\dmu W \Bigr]^{1/2}.\cr} \ref{\ualpbet}
$$
For the remaining integral over $W$, we have
$$
\int_\scd |w|^{2(m - |\gamma|-|\delta|)}   \Delta(w,w)^{-2s}
\dmu W  = \int_\scd |w|^{2(m - |\gamma| - |\delta|)}  d\mu_{r'}(W),  \num
$$
where $r'$ and $r$ differ by a constant.  We can apply
([6], Lemma 3.1~(ii))
to bound this expression by a constant times $r^{-m+|\gamma|+|\delta|}$.
Returning to \ualpbet, since $N(Z,Z) = \Delta(z,z) + ${\it nilpotent},
the components of $\rker ZZ^{1/2}$ can be bounded by
$\Delta(z,z)^{-r/2 - s''}$ for some $s''$ (the $s''$ occurs when
we Taylor expand $(\Delta + ${\it nilpotent}$)\inv$).
We thus have
$$
|u_{\alpha\beta}(z)| \le C r^{-m/2} \norm\psi\;
\norm{g}_t \;\Delta(z,z)^{-r/2 - s'- s''}.
\num
$$

Applying these results to \zintbnd, we find that
$$
|R| \le Cr^{-m/2} \norm g_t \>\norm f_0\> \norm\phi\> \norm\psi
\left|\int_\scd X_{S_f}(z)
\Delta(z,z)^{-r-2(s'+s'')} \dmu Z\right|^{1/2}. \ref{\rstep}
$$
The $\theta$ integration in the remaining integral
can be estimated using \fintprop:
$$
\eqalign{&\int_\scd X_{S_f}(z) \Delta(z,z)^{-r-2(s'+s'')} \dmu Z \cr
&\qquad= Cr^{n_1} \Lambda_r \int_{S_f} \Delta(z,z)^{- p_0 - 2(s'+s'')} dz
[1 + O(r\inv)].\cr} \num
$$
The integral over $S_f$ is finite and independent of $r$, so
we can absorb it into the constant.
According to \normprop, the
normalization constant $\Lambda_r$ can be bounded by a constant times
$r^{n_0 - n_1}$ as $r\to\infty$.
Applying all of this to \rstep, we have
$$
|R| \le C_{S_f}  r^{-(m - n_0)/2} \norm g_t \;\norm f_t
\;\norm\phi \;\norm\psi .
\num
$$
With the fact that $m - n_0 > 4$, this completes the
proof.  $\quad\square$

\vfill\eject

\noindent
{\bf References}
\baselineskip=12pt
\frenchspacing

\newcount\refnum
\global\refnum=0
\def\refitem{\global\advance\refnum by 1
	\medskip\item{\the\refnum.}}

\bigskip
\refitem
Berezin, F.A.: Quantization, {\it Math. USSR Izvestija} {\bf8} (1974),
1109--1165.

\refitem
Berezin, F.A.: Quantization in complex symmetric spaces, {\it Math. USSR
Izvestija} {\bf9} (1975), 341--379.

\refitem
Berezin, F.A.: {\it Introduction to Superanalysis}, D. Reidel Publ. Co.,
Dordrecht
(1987).

\refitem
Borthwick, D., Klimek, S., Lesniewski, A., and Rinaldi, M.: Super
Toeplitz operators and non-perturbative deformation quantization
of supermanifolds, {\it Comm. Math. Phys.}, to appear.

\refitem
Borthwick, D., Lesniewski, A., and Rinaldi, M.:
Lie superspheres, to appear.

\refitem
Borthwick, D., Lesniewski, A., and Upmeier, H.: Non-perturbative
deformation quantization of Cartan domains, {\it J. Funct. Anal.}
{\bf112}, to appear.

\refitem
Coburn, L.A.: Deformation estimates for the Berezin-Toeplitz quantization,
{\it Comm. Math. Phys.} {\bf 149} (1992), 415--424.

\refitem
Connes, A.: Non-commutative differential geometry, {\it Publ. Math. IHES}
{\bf62} (1985), 41--144.

\refitem
Glimm, J., and Jaffe, A.: {\it Quantum Physics}, Springer Verlag, Berlin,
Heidelberg, New York (1987).

\refitem
Helgason, S.: {\it Differential Geometry, Lie Groups, and Symmetric
Spaces},
Academic Press, New York/London (1978).

\refitem
Hernandez Ruiperez, D., and Munoz Masque, J.: Global variational
calculus on graded manifolds, I. Graded jet bundles, structure
$1$-form and graded infinitesimal contact transformations,
{\it J. Math. Pures et Appl.} {\bf 63} (1984), 283--309.

\refitem
Klimek, S., and Lesniewski, A.: Quantum Riemann surfaces, I. The
unit disc, {\it Comm. Math. Phys.} {\bf 146} (1992), 103--122.

\refitem
Kostant, B.: Graded manifolds, graded Lie theory and prequantization,
{\it Lecture Notes in Mathematics} {\bf 570}, Springer Verlag,
Berlin, Heidelberg, New York (1977).

\refitem
Kostant, B., and Sternberg, S.: Symplectic reduction, BRS cohomology,
and infinite-dimensional Clifford algebras, {\it Ann. Phys.} {\bf 176}
(1987), 49--113.

\refitem
Loos, O.: {\it Bounded Symmetric Domains and Jordan Pairs}, Univ. of
California, Irvine (1977).

\refitem
Manin, Yu.: {\it Gauge Field Theory and Complex Geometry}, Springer Verlag,
Berlin, Heidelberg, New York (1988).

\refitem
Rieffel, M.: Deformation quantization of Heisenberg manifolds, {\it
Comm. Math. Phys.} {\bf 122}, 531--562 (1989).

\refitem
Upmeier, H.: Toeplitz $\Bbb{C}^*$-algebras on bounded symmetric
domains,
{\it Ann. Math.} {\bf119} (1984), 549--576.

\vfill\eject\end